\documentclass[prx,aps,twocolumn,superscriptaddress,10pt]{revtex4-2}
\usepackage[utf8]{inputenc}
\usepackage[T1]{fontenc}
\usepackage{color}
\usepackage{amsmath}
\usepackage{amsfonts}
\usepackage{hyperref}
\usepackage{amsthm}
\usepackage{bm}
\usepackage{float}
\usepackage{textcomp}
\usepackage{upgreek}
\usepackage{textgreek}
\usepackage{setspace}
\usepackage{amssymb,graphicx}
\usepackage[percent]{overpic}
\hypersetup{
  colorlinks = true,
  allcolors= blue,
}
\usepackage{tikz,pgfplots}
\usepackage{blkarray}

\usetikzlibrary{patterns}
\usetikzlibrary{fadings}
\usetikzlibrary{decorations.pathmorphing,patterns}
\tikzset{snake it/.style={decorate, decoration=snake}}
\usetikzlibrary{arrows, decorations.markings}
\pgfplotsset{compat=1.14}
\tikzset{
vecArrow/.style={
  thick,
  decoration={markings,mark=at position
   1 with {\arrow[scale=2,thin]{open triangle 60}}},
  double distance=1.4pt, shorten >= 10.5pt,
  preaction = {decorate},
  postaction = {draw,line width=1.4pt, white,shorten >= 4.5pt}
  },
innerWhite/.style={
  semithick,
  white,
  line width=1.4pt,
  shorten >= 4.5pt
  }
}
\definecolor{orange}{rgb}{1,0.5,0}
\definecolor{darkgreen}{rgb}{0,0.4,0.1}

\newcommand{\WidthFigure}{\columnwidth}

\definecolor{redMC}{rgb}{0.568,0.1254,0.0705}
\definecolor{bluesilica}{rgb}{0.1254,0.3058,0.6117}
\definecolor{cyansilica}{rgb}{0.4588,0.6470,0.9215} 

\makeatletter
\newcommand{\miniscule}{\@setfontsize\miniscule{4}{5}}
\makeatother

\makeatletter
\newcommand{\doublehatSub}[1]{%
\begingroup%
  \let\macc@kerna\z@%
  \let\macc@kernb\z@%
  \hat{\raisebox{-.07ex}{\vphantom{\ensuremath{#1}}}\smash{\hat{#1}}}%
\endgroup%
}
\makeatother

\makeatletter
\DeclareFontFamily{OMX}{MnSymbolE}{}
\DeclareSymbolFont{MnLargeSymbols}{OMX}{MnSymbolE}{m}{n}
\SetSymbolFont{MnLargeSymbols}{bold}{OMX}{MnSymbolE}{b}{n}
\DeclareFontShape{OMX}{MnSymbolE}{m}{n}{
    <-6>  MnSymbolE5
   <6-7>  MnSymbolE6
   <7-8>  MnSymbolE7
   <8-9>  MnSymbolE8
   <9-10> MnSymbolE9
  <10-12> MnSymbolE10
  <12->   MnSymbolE12
}{}
\DeclareFontShape{OMX}{MnSymbolE}{b}{n}{
    <-6>  MnSymbolE-Bold5
   <6-7>  MnSymbolE-Bold6
   <7-8>  MnSymbolE-Bold7
   <8-9>  MnSymbolE-Bold8
   <9-10> MnSymbolE-Bold9
  <10-12> MnSymbolE-Bold10
  <12->   MnSymbolE-Bold12
}{}

\let\llangle\@undefined
\let\rrangle\@undefined
\DeclareMathDelimiter{\llangle}{\mathopen}%
                     {MnLargeSymbols}{'164}{MnLargeSymbols}{'164}
\DeclareMathDelimiter{\rrangle}{\mathclose}%
                     {MnLargeSymbols}{'171}{MnLargeSymbols}{'171}
\makeatother
\DeclareUnicodeCharacter{300}{à}

\DeclareMathAlphabet{\mathsfit}{\encodingdefault}{\sfdefault}{m}{sl}
\SetMathAlphabet{\mathsfit}{bold}{\encodingdefault}{\sfdefault}{bx}{sl}
\newcommand{\tens}[1]{\bm{\mathsfit{#1}}}
\newcommand{\tenscomp}[1]{\mathsfit{#1}}

\newcommand{\bDiamond}{\mathbin{\Diamond}}

\makeatletter
\newcommand\bigDiamond{\mathop{\mathpalette\bigDi@mond\relax}}
\newcommand\bigDi@mond[2]{%
  \vcenter{\hbox{\m@th
    \scalebox{\ifx#1\displaystyle 2\else1.2\fi}{$#1\Diamond$}%
  }}%
}
\newcommand\bigLozenge{\mathop{\mathpalette\bigL@zenge\relax}}
\newcommand\bigL@zenge[2]{%
  \vcenter{\hbox{\m@th
    \scalebox{\ifx#1\displaystyle 2\else1.2\fi}{$#1\blacklozenge$}%
  }}%
}
\makeatother
\makeatletter
\usepackage{comment}
\let\wfs@comment@comment\comment
\let\comment\@undefined

\usepackage{changes}
\let\wfs@changes@comment\comment
\let\comment\@undefined

\newcommand\comment{%
    \ifthenelse{\equal{\@currenvir}{comment}}
    {\wfs@comment@comment}
    {\wfs@changes@comment}%
}
\makeatother

\makeatother

\definecolor{dgreen}{rgb}{0,0.45,0}

\makeatletter
\@namedef{Changes@AuthorColor}{red}
\colorlet{Changes@Color}{red}
\setaddedmarkup{\textcolor{dgreen}{#1}}
\setdeletedmarkup{\textcolor{red}{#1}}
\makeatother
\begin{document}

\setcitestyle{super}

\title{\vspace*{-5mm}Temperature-invariant heat conductivity from compensating crystalline and glassy transport: from the Steinbach meteorite to furnace bricks
\vspace*{-2mm}}

\author{Michele Simoncelli}
\email{ms2855@cam.ac.uk}
\affiliation{ 
Theory of Condensed Matter Group of the Cavendish Laboratory, University of Cambridge (UK)}

\author{Daniele Fournier}
\affiliation{Sorbonne Université, CNRS, Institut des NanoSciences de Paris, INSP, UMR7588, F-75005 Paris, France}

\author{Massimiliano Marangolo}
\affiliation{Sorbonne Université, CNRS, Institut des NanoSciences de Paris, INSP, UMR7588, F-75005 Paris, France}

\author{Etienne Balan}
\affiliation{Sorbonne Université, CNRS, Institut de Minéralogie, de Physique des Matériaux et de Cosmochimie, 75252 Paris, France}

\author{Keevin Béneut}
\affiliation{Sorbonne Université, CNRS, Institut de Minéralogie, de Physique des Matériaux et de Cosmochimie, 75252 Paris, France}

\author{Benoit Baptiste}
\affiliation{Sorbonne Université, CNRS, Institut de Minéralogie, de Physique des Matériaux et de Cosmochimie, 75252 Paris, France}

\author{Béatrice Doisneau}
\affiliation{Sorbonne Université, CNRS, Institut de Minéralogie, de Physique des Matériaux et de Cosmochimie, 75252 Paris, France}

\author{Nicola Marzari}
\affiliation{Theory and Simulation of Materials, and National Centre for Computational Design and Discovery of Novel Materials, {\'E}cole Polytechnique F{\'e}d{\'e}rale de Lausanne, Lausanne, Switzerland.}

\author{Francesco Mauri}
\affiliation{Dipartimento di Fisica, Universit{\`a} di Roma La Sapienza, Italy}

\begin{abstract}
The thermal conductivities of crystals and glasses vary strongly and with opposite trends upon heating, decreasing in crystals and increasing in glasses. 
Here, we show---both with first-principles predictions based on the Wigner transport equation and with thermoreflectance experiments---that the dominant transport mechanisms of crystals (particle-like propagation) and glasses (wave-like tunnelling) can coexist and compensate in materials with crystalline bond order and nearly glassy bond geometry. 
We demonstrate that ideal compensation emerges in a sample of silica in the form of tridymite, carved from a meteorite found in Steinbach (Germany) in 1724, and yields a `propagation-tunneling-invariant' (PTI) conductivity that is independent from temperature and intermediate between the opposite trends of $\alpha$-quartz crystal and silica glass. 
We show how such PTI conductivity occurs in the quantum regime below the Debye temperature, 
and can largely persist at high temperatures in a geometrically amorphous tridymite phase found in refractory bricks fired for years in furnaces for steel smelting. Last, we discuss implications to heat transfer in solids exposed to extreme temperature variations, ranging from planetary cooling to heating protocols to reduce the carbon footprint of industrial furnaces.    
\end{abstract}

\maketitle

The relation between atomistic structure and macroscopic thermal conductivity of insulating solids is critical in many and diverse scientific problems,  from sustainable energy technologies\cite{kimber_dynamic_2023,yan_high-performance_2022,selezneva_strong_2021} to the formation of meteoroids\cite{murphy_quinlan_conductive_2021} and planet differentiation \cite{whittington_temperature-dependent_2009}.
The first insights onto such relation were provided by \citet{peierls1929kinetischen} for ordered crystals, where anharmonicity is the main source of thermal resistance. In these cases structural order allows for the emergence of atomic vibrational excitations (phonon wavepackets) with dispersions and lifetimes determined by interatomic forces, and a particle-like dynamics ruled by the phonon Boltzmann equation (BTE).
The BTE universally predicts a strong decrease of the conductivity with temperature ($\kappa(T){\sim} T^{-1}$ for dominant three-phonon interactions and $T$ larger than the Debye temperature $T_D$), which accurately reproduces several experiments in clean crystals \cite{broido_intrinsic_2007,qian_phonon-engineered_2021}.
In contrast, amorphous solids experimentally display a $\kappa(T)$ increasing with temperature up to saturation\cite{cahill_lower_1992}. Insights onto such saturation were obtained 
by Cahill, Watson and Pohl (CWP) \cite{cahill_lower_1992}, who argued that an early conduction model proposed by Einstein \cite{Einstein_1911}---describing heat transport as a random walk occurring between partially localized and uncorrelated atomic oscillators---could be used to realistically describe disordered solids. CWP refined such model, focusing on solids where vibrations can be approximatively described as particle-like excitations with dispersion and damping due to disorder only, predicting a universal trend for $\kappa(T)$ increasing up to saturation. An analogous trend was predicted by \citet{allen1989thermal} (AF) when disorder destroys vibrations' dispersion, in this case as a consequence of wave-like tunneling mechanisms between quasi-degenerate vibrational eigenstates.
Recently, the BTE and the AF formulation have been unified into a general Wigner transport equation (WTE)\cite{simoncelli2019unified,simoncelli_wigner_2022}, which accounts for both anharmonicity and disorder. As a result, the WTE generalizes and extends BTE and AF,
encompassing the emergence and coexistence of temperature-inhibited propagation and temperature-activated tunneling transport mechanisms, allowing  hybrid
trends for $\kappa(T)$. 
Here we report a material displaying such hybrid trend, where the two conductivity components compensate in a  `propagation-tunneling-invariant' (PTI) conductivity.
We discuss the connection between macroscopic PTI conductivity and atomic structure, as well as implications to heat management in industrial furnaces and possible PTI signatures in planetary science.

\noindent
\textbf{Theory of PTI conductivity}\\
To elucidate the fundamental origin of PTI conductivity, we start from the 
WTE conductivity expressed as\cite{simoncelli_wigner_2022},
\begin{equation}
\begin{split}
&\kappa^{\alpha\beta}{=}\kappa^{\alpha\beta}_P{+}\frac{1}{\mathcal{V}{N_{\rm c}} }{\sum_{\bm{q},s\neq s'}\hspace*{-1.5mm}} 
\frac{\omega(\bm{q})_s{+}\omega(\bm{q})_{s'}}{4}\!\left[\frac{C(\bm{q})_{s}}{\omega(\bm{q})_{s}}{+}\frac{C(\bm{q})_{s'\!}}{\omega(\bm{q})_{s'\!}}\right]\!\!{\times}\\
&\tenscomp{v}^{\alpha}(\bm{q})_{s,s'}\tenscomp{v}^{\beta}(\bm{q})_{s',s}
\frac{\frac{1}{2}\big[\Gamma(\bm{q})_s{+}\Gamma(\bm{q})_{s'}\big]}{\big[\omega(\bm{q})_s{-}\omega(\bm{q})_{s'}\big]^2+\frac{1}{4}\big[\Gamma(\bm{q})_s{+}\Gamma(\bm{q})_{s'}\big]^2}
\;;
\label{eq:thermal_conductivity_combined}
\end{split}
\raisetag{13mm}
\end{equation}
where $\hbar\omega(\bm{q})_s$ is the energy of a vibration having wavevector $\bm{q}$ and mode $s$, and carrying the specific heat 
\begin{equation}
C(\bm{q})_s{=}C[\omega(\bm{q})_s,T]{=}\tfrac{\hbar^2\omega^2(\bm{q})_s }{k_{\rm B} T^2} \bar{\tenscomp{N}}(\bm{q})_s\big[\bar{\tenscomp{N}}(\bm{q})_s{+}1\big],  
\label{eq:quantum_specific_heat_A}
\end{equation}
with $\bar{\tenscomp{N}}(\bm{q})_s{=}[\exp(\hbar \omega(\bm{q})_s/k_{\rm B}T){-}1]^{-1}$ being the Bose-Einstein distribution at temperature $T$.
$\tenscomp{v}^{\alpha}(\bm{q})_{s,s'}$ is the velocity operator coupling vibrational modes  $s$ and $s'$ at the same wavevector $\bm{q}$ ($\alpha$ is a Cartesian direction) \cite{simoncelli_wigner_2022}, $N_{\rm c}$ is the number of $\bm{q}$ in the sum and $\mathcal{V}$ the primitive-cell volume; $\hbar\Gamma(\bm{q})_s$ is the vibration's linewidth, which accounts for 
anharmonicity and isotopic disorder\cite{simoncelli_wigner_2022}. 
It can be shown\cite{simoncelli_wigner_2022,fiorentino_green-kubo_2023} that $\kappa^{\alpha\beta}_P$ coincides with the BTE conductivity\cite{peierls1929kinetischen}, which describes particle-like heat transport.   
This is apparent from its form in the relaxation-time approximation (employed here, and accurate in SiO$_2$ polymorphs\cite{simoncelli_thermal_2023}):
$\kappa_P^{\alpha\beta}=\frac{1}{\mathcal{V} N_C }\sum_{\bm{q}s}C[\omega(\bm{q})_{s},T]{\tenscomp{v}^{\alpha}(\bm{q})_{s,s}}\Lambda^\beta(\bm{q})_s$ describes particle-like excitations having energy $\hbar\omega(\bm{q})_{s}$, and propagating with velocity ${\tenscomp{v}^{\alpha}(\bm{q})_{s,s}}$ over the mean free path $\Lambda^\beta(\bm{q})_{s}={\tenscomp{v}^{\beta}(\bm{q})_{s,s}}[\Gamma(\bm{q})_s]^{-1}$.  
The other term in Eq.~(\ref{eq:thermal_conductivity_combined}) accounts for wave-like tunneling of heat between pairs of vibrations having energy difference smaller than the linewidths, and is referred to as coherences conductivity\cite{simoncelli2019unified}, $\kappa_C^{\alpha\beta}$.   

In `simple' crystals with phonon interband spacings larger than the linewidths ($|\omega(\bm{q})_s{-}\omega(\bm{q})_{s'}|\gg\frac{1}{2}\big[\Gamma(\bm{q})_s{+}\Gamma(\bm{q})_{s'}\big]$ $\forall s{\neq}s'$) one can show that 
$\kappa_P^{\alpha\beta}{\gg}\kappa_C^{\alpha\beta}$, and thus the WTE conductivity~(\ref{eq:thermal_conductivity_combined}) reduces to the BTE one.
In the opposite amorphous limit, disorder destroys the phonon dispersion (the lack of periodicity implies a diverging cell volume $\mathcal{V}$, thus only $\bm{q}{=}\bm{0}$ contributes to Eq.~(\ref{eq:thermal_conductivity_combined})) and forbids perfect degeneracies between vibrational energy levels. This implies the emergence of quasi-degenerate vibrational states ($|\omega(\bm{0})_s{-}\omega(\bm{0})_{s'}|{\ll}\frac{1}{2}\big[\Gamma(\bm{0})_s{+}\Gamma(\bm{0})_{s'}\big], s{\neq} s'$), which can be coupled by off-diagonal velocity-operator elements and contribute to  $\kappa_C^{\alpha\beta}$. 
We also note that in the harmonic disordered limit $\hbar\Gamma(\bm{0})_s{\to}\hbar\eta{\to} 0\;\forall s$ these wave-like mechanisms reduce to those described by the AF theory\cite{allen1989thermal,simoncelli_thermal_2023}. 
\begin{figure}[b]
\vspace*{-2mm}
  \includegraphics[width=0.9\WidthFigure]{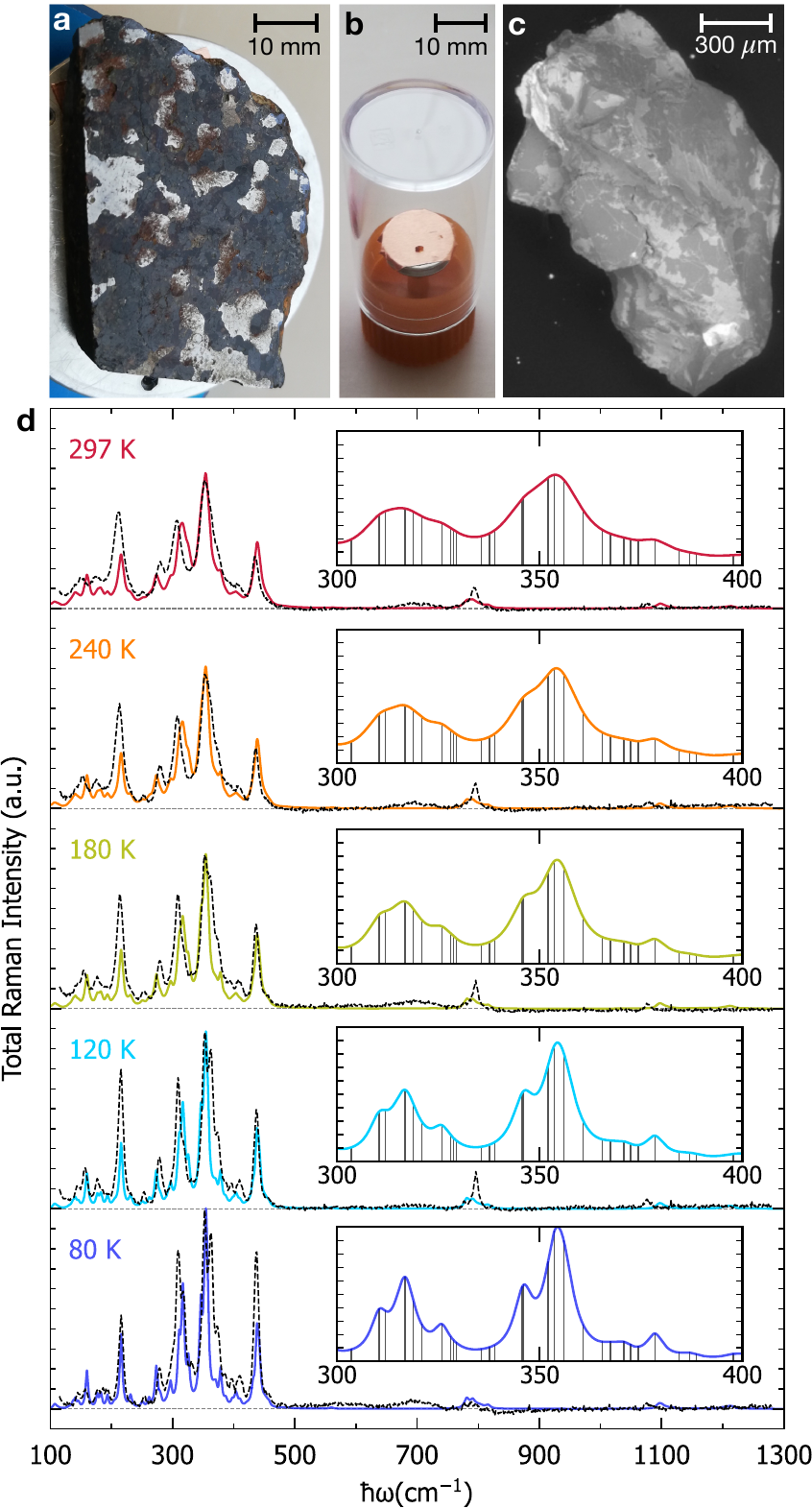}\vspace*{-3mm}
  \caption{\textbf{Raman spectra of MC tridymite.}
   We isolated a MC tridymite grain from a fragment of the Steinbach meteorite (Karl-Marx-Stadt, Germany, 1724, panel \textbf{a}),  
  \textbf{b} shows the optical view, and \textbf{c} the SEM view. The experimental Raman spectra (\textbf{d}, dashed) agrees with simulations (\textbf{d}, solid); the insets show individual Raman-active modes as vertical lines.}
  \label{fig:raman}
\end{figure}
\begin{figure*}
\vspace*{-5mm}
  \includegraphics[width=\textwidth]{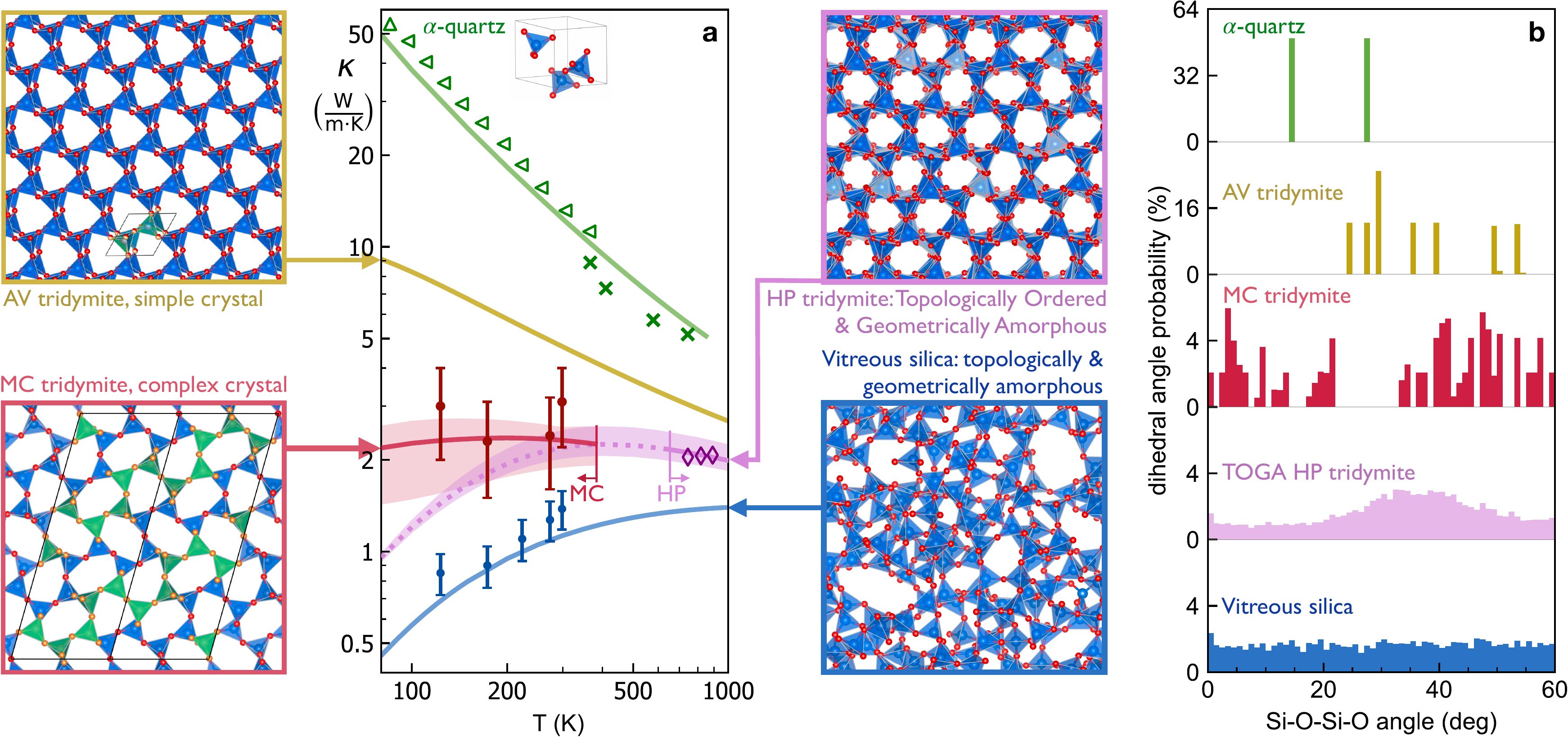}\vspace*{-3mm}
  \caption{\textbf{Thermal conductivity of SiO$_2$ polymorphs with various degree of disorder.}
   Panel  \textbf{a}, lines are predictions from the Wigner transport equation, scatter points are experiments. Green (\textcolor{darkgreen}{$\bm\lhd$} \cite{Eucken1911}, 
   \textcolor{darkgreen}{\textsf{x}}\cite{Knapp1943}) and yellow are the simple crystals $\alpha-$quartz (stable up to 847 K\cite{richet1982thermodynamic}) and average (AV) tridymite, respectively, both along the $z$ direction. Red is MC tridymite (stable up to 383 K\cite{graetsch_modulated_2009}) along the direction in which our conductivity measurements (\textcolor{redMC}{${\bullet}\hspace*{-1.45mm}\rm{I}$}) intercept the largest crystalline domain of the sample (see text);
   the shaded area shows the maximum and minimum eigenvalues of the conductivity tensor. In the yellow and red panels, the black lines show the boundaries of the primitive cell of AV and MC tridymite, respectively, and the SiO$_4$ tetrahedra inside them are highlighted in green.
   Pink is Topologically Ordered \& Geometrically Amorphous (TOGA) tridymite (\textcolor{purple}{\textcolor{purple}{$\bm\bDiamond$}}\cite{Pierce_1936}). This phase is found above 653 K\cite{graetsch_modulated_2009}, it features only 12-atoms rings (as AV tridymite) and bond geometries distorted around an ideal hexagonal-crystal phase (HP). 
   Blue is completely disordered isotropic vitreous silica (our measurements are \textcolor{bluesilica}{${\bullet}\hspace*{-1.45mm}\rm{I}$}).
   Panel \textbf{b,} the spread of the Si-O-Si-O dihedral angle distribution quantifies geometric disorder. 
   }
  \label{fig:k_main_text}
\end{figure*}
Most importantly, Eq.~(\ref{eq:thermal_conductivity_combined}) in principle covers all the hybrid cases, where particle-like and wave-like mechanisms coexist and have comparable strength; these are also labeled as `complex crystals'~\cite{simoncelli2019unified} and feature many vibrational modes $s,s'$ at wavevector $\bm{q}$ having energy difference comparable to the linewidths, \textit{i.e.} $|\omega(\bm{q})_s{-}\omega(\bm{q})_{s'}|{\sim}\frac{1}{2}\big[\Gamma(\bm{q})_s{+}\Gamma(\bm{q})_{s'}\big]$ \cite{simoncelli_wigner_2022}. 
Eq.~(\ref{eq:thermal_conductivity_combined}) suggests that the appearance of such hybrid  regime requires a certain degree of disorder or anharmonicity. 
The former is temperature-independent and determines the average energy-level spacings \cite{simoncelli_wigner_2022}; the latter, instead, depends on temperature, and is mainly determined by the chemistry \cite{di_lucente_crossover_2023}. 
Hybrid behavior at high temperature ($T{>}T_D$) has been discussed for highly anharmonic materials\cite{simoncelli2019unified,di_lucente_crossover_2023,tadano2021firstprinciples,Anees_2023,simoncelli_wigner_2022}, whose  conductivity is generally low (${\lesssim} 3$ W/mK) and varies weakly (less than a factor of two) between different structural phases \cite{tadano2021firstprinciples,simoncelli_wigner_2022}.
In contrast, the possibility to drive hybrid behavior exploiting structural disorder in weakly anharmonic materials with well-defined phonon bands has hitherto not been explored, probably due to challenges in controlling complex atomic structures.
However, these case are promising for engineering materials with target conductivities, since the conductivities of ordered and disordered polymorphs can differ of several orders of magnitude. E.g., in silicon dioxide (SiO$_2$) crystalline $\alpha$-quartz and vitreous silica have similar  anharmonicity, but their conductivities have opposite trends in temperature  and their magnitude at 80 K differs of about a factor 100 \cite{simoncelli_thermal_2023}.
While these paradigmatic cases have been extensively studied, the conductivities of hybrid polymorphs with intermediate degree of disorder have not received attention, due to rarity of samples and lack of theoretical methods capable to describe their conductivity below $T_D$\cite{Puligheddu2019,simoncelli_thermal_2023}. 
Thus, we analyze them now using the WTE and experiments.

\noindent
\textbf{Proof of concept of PTI: meteoritic tridymite}\\
Monoclinic (MC) tridymite is an hybrid SiO$_2$ polymorph found in extraterrestrial bodies such as the Steinbach meteorite\cite{Dollase76} (Fig.~\ref{fig:raman}, \textbf{a}, \textbf{b}, \textbf{c}), Mars\cite{morris_silicic_2016}, and planetary embryos\cite{connelly_pb_2019}; interestingly, it was also found on the surface of refractory silica bricks fired in coke ovens for several years \cite{hirose_x-ray_2005}.
 It combines order in the bond topology and nearly glassy bond geometry: its primitive cell 
 contains 48 oxygen and 24 silicon atoms, covalently bonded into a structure with reduced number of crystal symmetries ($Cc$ spacegroup\cite{graetsch_modulated_2009}, MC tridymite is a commensurate superstructure of an average `AV tridymite' phase, a geometrically ordered structure that contains 12 atoms per primitive cell).
The large periodicity length and related large number of symmetry-inequivalent atoms in the primitive cell of MC tridymite translates into a large number of Raman-active vibrational modes.
The Raman spectrum, $I(\omega) {=} \sum_{s} I_s \frac{\Gamma_s/2}{(\omega-\omega_s)^2+(\Gamma_s/2)^2}$ (where  $I_s$, $\hbar\omega_s$, and $\hbar\Gamma_s$ are the Raman activity, energy, and full linewidth of mode $s$ at $\bm{q}{=}\bm{0}$, respectively, see Methods), provides information on the overlap and wave-like couplings between the different modes. A  sharp (smooth) spectrum indicates negligible (strong) overlap between modes; suggesting simple-crystal (complex-crystal) behavior in heat transport. 
In Fig.~\ref{fig:raman}\textbf{d} we compare the theoretical and experimental Raman spectra of MC tridymite. 
The sample was extracted from the stony-iron (IVA) Steinbach meteorite, and its identification with MC tridymite is based on the agreement between Raman theoretical predictions and experiments. 
The insets highlight that the number of distinguishable peaks in the Raman spectrum decreases upon increasing temperature, a behavior originating from the increase of the anharmonic linewidths with temperature.
In fact, the equation for $I(\omega)$ shows that, to distinguish peaks of different modes, linewidths have to be smaller than their energy difference. This occurs at low temperature, but increasing temperature the linewidths increase, yielding overlaps between modes that smoothen the spectrum. A smooth Raman spectrum with overlaps suggests 
coexistence of particle-like and wave-like transport \cite{simoncelli_wigner_2022} within the WTE.

\begin{figure*}
\vspace*{-5mm}
  \includegraphics[width=\textwidth]{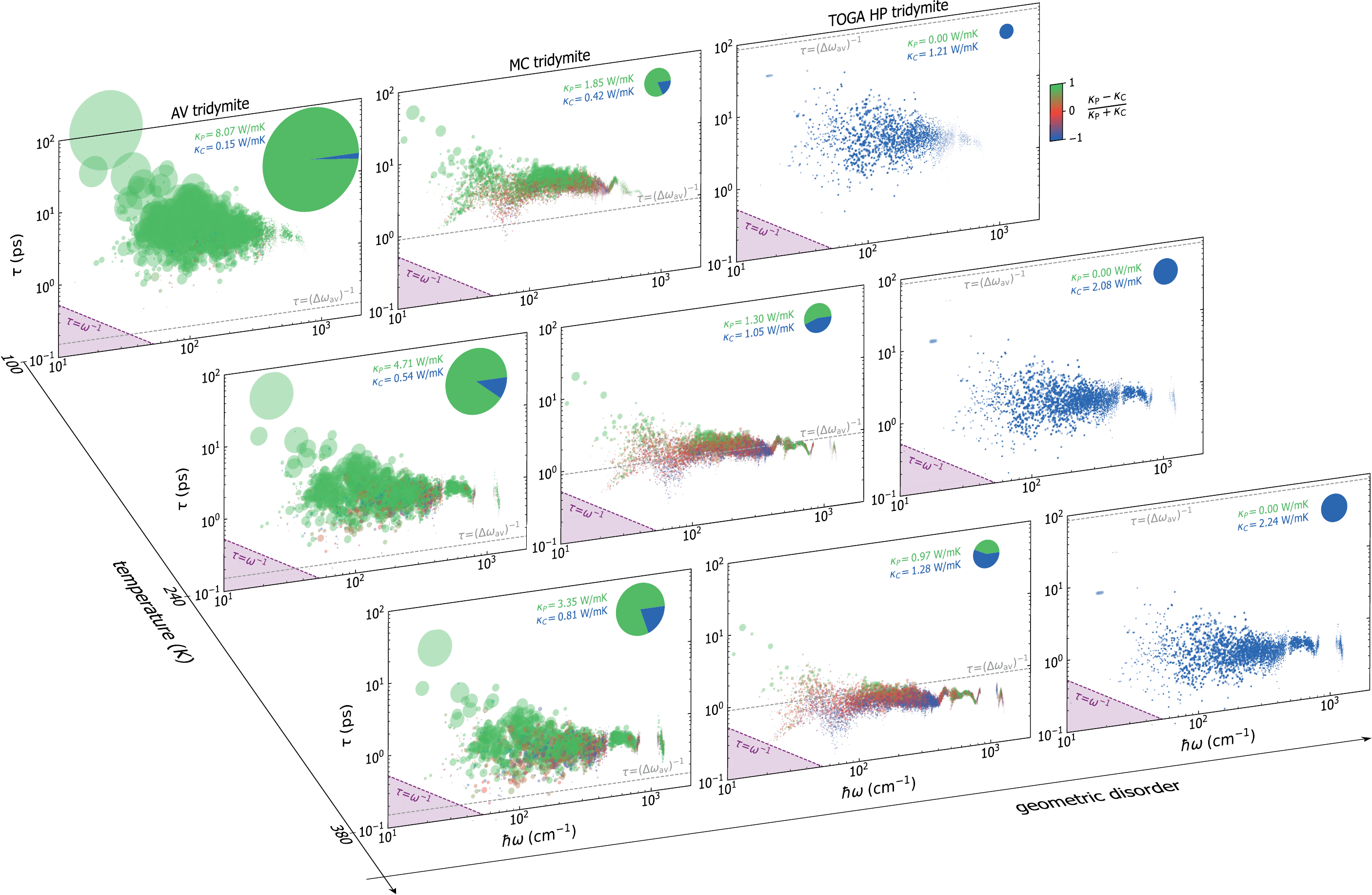}\\[-4mm]
  \caption{\textbf{Particle-like propagation and wave-like tunneling in polymorphs with various geometric disorder,} which
 increases from left to right. Temperature increases from top to bottom. 
 We show the lifetimes as a function of energy for each vibrational mode, represented with a circle having area proportional to the contribution to conduction and color related to the type of contribution. Green and blue represents particle-like and wave-like mechanisms, respectively, and intermediate colors show coexistence of the two (e.g., red is 50 $\%$ of each). The horizontal gray-dashed line is the `Wigner limit in time', a timescale that increases with structural complexity and determines the center of the particle-wave crossover\cite{simoncelli_wigner_2022}.
The pie charts have a radius proportional to the total conductivity, with green and blue slices representing the propagation and tunneling contributions, respectively.
The purple region is the overdamped regime, where vibrations are no longer well defined quasiparticle excitations \cite{simoncelli_wigner_2022}.}
  \label{fig:lifetime}
\end{figure*}

These expectations are quantitatively confirmed in Fig.~\ref{fig:k_main_text}: predictions and experiments (panel \textbf{a}) demonstrate that the thermal conductivity of MC tridymite combines the decreasing trend of simple crystals such as $\alpha$-quartz or AV tridymite (which feature order in both the bond topology and geometry), and the increasing trend of vitreous silica (disordered in both topology and geometry).  
We show in Fig.~\ref{fig:balancing} in the Methods that the temperature-invariant thermal conductivity originates from compensating variations of the particle-like and wave-like conductivities, and this occurs at temperatures (80-380 K) where
both these contributions are influenced by the quantum, Bose-Einstein statistics of vibrations---to confirm this we also show in Fig.~\ref{fig:spec_heat} that the specific heat at these temperatures is far from the classical limit.
Thus, the mechanism driving conductivity invariance here is fundamentally different from the AF or CWP conductivity saturation, which both follow from the classical saturation of the specific heat\cite{allen1989thermal,cahill_lower_1992,xia_unified_2023}; as such we will hereafter refer to it  as `propagation-tunneling-invariant' (PTI) conductivity. 
The red line in Fig.~\ref{fig:k_main_text} shows the prediction for the conductivity along the direction in which the thermoreflectance conductivity measurement intercepts the largest crystalline domain of the sample, \textit{i.e.} the vector $\bm{m}{=}0.65 \bm{a}{-}4.7\bm{b}{+}0.715\bm{c}$, where $\bm{a}$, $\bm{b}$, $\bm{c}$ are the lattice vectors of MC tridymite\cite{Dollase76} (see Methods for details); the shaded red area ranges from the minimum to the maximum eigenvalues of the conductivity tensor.  
Most importantly, we see that around 380 K---the temperature at which MC tridymite undergoes a phase transition\cite{graetsch_modulated_2009}---the conductivity of MC tridymite coincides with that of a silica phase characterized by crystal-like bond topology (equivalent to AV tridymite) and glass-like bond geometry (close to that of vitreous silica, as shown in panel Fig.~\ref{fig:k_main_text}\textbf{b}). 
We show in the Methods how such topologically ordered and geometrically amorphous (TOGA) phase features only 12-atom rings (as AV tridymite), 
and its bond angles are distorted (by thermally-induced fluctuations) around those of a 
highly symmetric, crystalline hexagonal phase (HP) known as HP tridymite\cite{graetsch_modulated_2009}.
Our findings agree with those by \citet{graetsch_modulated_2009}, who states that above 653 K tridymite features bond angles that are distorted around those of HP tridymite.
TOGA HP tridymite and vitreous silica have similar density and specific heat (see Methods);  
their thermal conductivities have similar increasing trend but different magnitude, with TOGA HP tridymite having the largest conductivity. 
This difference can be related to their different disorder in the bond topology, present in vitreous silica and absent in TOGA HP tridymite. Specifically, we show in the Methods that higher topological disorder yields smaller velocity-operator elements and thus smaller conductivity.
Importantly, Refs.~\cite{curtis2010postmortem,wereszczak_postmortem_2000}, show that the `postmortem' composition of refractory silica bricks fired for years in furnaces\cite{kasaihistory} is mainly tridymite in the `cold-end' region ($700\; K\lesssim T \lesssim 1000\; K$), \textit{i.e.} exposed to temperatures in the stability range of TOGA HP tridymite  \cite{graetsch_modulated_2009}.
As such, these insights on the conductivity difference between TOGA HP tridymite and vitreous silica may be exploited to
increase the efficiency steel furnaces, where a conductivity increase may lead to energy saving\cite{yang_performance_nodate} and consequently reduction of CO$_2$ emissions\cite{CO2_furnace}.\\[2mm]
\noindent
\textbf{Microscopic origin of PTI conductivity}\\
We now focus on understanding the microscopic physics that determines the differences in the macroscopic thermal conductivities of polymorphs having the same bond topology but different geometry. To gain insights, we resolve in Fig.~\ref{fig:lifetime} the conductivities of AV tridymite, MC tridymite, and TOGA HP tridymite  
in terms of microscopic contributions from individual vibrational modes.
Specifically, for each mode $(\bm{q})_s$ we show the lifetime
$\tau(\bm{q})_s=[\Gamma(\bm{q})_s]^{-1}$ as a function of energy $\hbar\omega(\bm{q})_s$, also quantifying how much it contributes to the particle-like and wave-like conductivities (see Methods for details).
Focusing on the simple crystal AV tridymite, we see that its large conductivity is mainly determined at all temperatures by  particle-like vibrations.
Such behavior originates from having, at all temperatures, negligible overlaps between different vibrational eigenstates (phonon bands), corresponding to vibrations' lifetimes much larger than the `Wigner' limit in time---we recall that the latter is a timescale that determines the center of the particle-wave crossover for vibrations; it is inversely related to the average energy-level spacing,  thus it increases with structural complexity\cite{simoncelli_wigner_2022}.
Turning our attention to MC tridymite, we see that its higher structural complexity yields a larger Wigner limit in time, and many vibrational modes cross such limit as the anharmonic lifetimes decrease with temperature (this corresponds to having linewidths that become larger than the average energy-level spacing).
As a result, the particle-like conductivity decreases and the wave-like conductivity increases; as discussed before, these variations compensate and determine the constant, PTI conductivity.  
Finally, in TOGA HP tridymite the strong geometric disorder forbids degeneracies between vibrational modes, yielding a very large Wigner limit in time and consequently a conductivity entirely determined by wave-like mechanisms and increasing with temperature.
We highlight how the wave-like conductivity of TOGA HP tridymite is larger than the wave-like conductivity of MC tridymite. 
To understand the microscopic origin of this difference, we recall that the strength of wave-like transport is determined by the magnitude of off-diagonal velocity-operator elements (see Eq.~(\ref{eq:thermal_conductivity_combined}), and we show in the Methods (Fig.~\ref{fig:velop}) that these elements become larger upon increasing geometric disorder. To confirm that such increase of wave-like conductivity is driven mainly by disorder, we also show that the specific heat displays negligible variations upon increasing disorder (Fig.~\ref{fig:spec_heat}), and anharmonicity becomes less important as disorder increases (Fig.~\ref{fig:eff_anh}).
We note that the wave-like conductivities of TOGA HP tridymite and MC tridymite have a similar trend (pink line in Fig.~\ref{fig:k_main_text} and blue line in Fig.~\ref{fig:balancing}); moreover, increasing temperature the particle-like conductivity of MC tridymite decreases (Fig.~\ref{fig:balancing}); this  implies that at very high temperature (above 650 K) MC tridymite would have a conductivity mainly determined by wave-like transport.
As such, above 650 K we interpret the nearly temperature-invariant (almost saturated) wave-like-only conductivity of TOGA HP tridymite as a reminiscence of the wave-like conductivity of MC tridymite.\\[2mm]
\noindent
\textbf{Conclusions and outlook}\\
We have shown that meteoritic MC tridymite
is a proof-of-concept material in which `propagation-tunneling-invariant' (PTI) transport occurs, since it features a temperature-invariant conductivity due to the coexistence and compensation of crystalline (particle-like propagation) and glassy (wave-like tunneling) heat-transport mechanisms.
We have shown that the PTI mechanism can occur in the quantum, below-Debye-temperature regime in materials with complex but well defined crystal structure. 
As such, these insights extends our understanding on the physics underlying temperature-invariant  conductivity beyond the established classical saturation mechanisms discussed by Einstein (CWP\cite{cahill_lower_1992}) or Allen and Feldman\cite{allen1989thermal}, which are limited to high temperature and to solids with broken translational symmetry.
Importantly, we have shown that a reminiscence of the PTI mechanism persists in 
topologically ordered \& geometrically amorphous HP tridymite above 650 K, yielding a nearly temperature-invariant conductivity that, due to lack of disorder in the bond topology, is higher than the conductivity of fully amorphous vitreous silica. 
These insights on how removing bond-topology disorder drives an increase in the conductivity may be exploited to increase the efficiency of steel furnaces. Specifically, increasing the refractories' conductivity by 0.5 W/m K would reduce the burning period of hot blast stove of 8.2\%\cite{yang_performance_nodate}; considering that as of today producing 1 kg of steel yields about 1.4 kg of C0$_2$\cite{CO2_furnace}, these findings may directly contribute to global decarbonization efforts. 
{Moreover, the present insight 
could find direct applications also in planetary sciences. Specifically, the linear (non-linear) cooling dynamics of a planet(esimal) is related to the presence (absence) of PTI materials\cite{murphy_quinlan_conductive_2021} allowing to obtain information on the atomistic structure of celestial bodies from their cooling dynamics, and enabling to develop realistic geophysical models for, e.g., crust formation\cite{murphy_quinlan_conductive_2021} and differentiation\cite{whittington_temperature-dependent_2009,tosi_mantle_2013}. } Finally, we note that the PTI mechanism discussed here shares fundamental common underpinnings with other quasiparticle’s transport phenomena in solids---involving, e.g., electrons\cite{chakraborty_boltzmann_1979,cepellotti_interband_2021} or magnons\cite{wei_giant_2022}---thus our findings will potentially inspire developments and applications of PTI transport in multiple areas of materials science.


%


\begin{center}
\vspace{10mm}
  \textbf{\large METHODS}
\end{center}

\section{Influence of disorder on heat-transport mechanisms} 
\label{sec:relationship_between_structural_disorder_heat_transport_mechanisms_and_m}

In this section we provide details on how the different degree of structural disorder in the silica polymorphs shown in Fig.~\ref{fig:k_main_text} affects the microscopic vibrational properties and consequently the macroscopic conductivity~(\ref{eq:thermal_conductivity_combined}).

\subsection{Geometric disorder in SiO$_2$ polymorphs} 
\label{sub:geometric_disorder}
\begin{figure}[h]
  \centering
  \includegraphics[width=\WidthFigure]{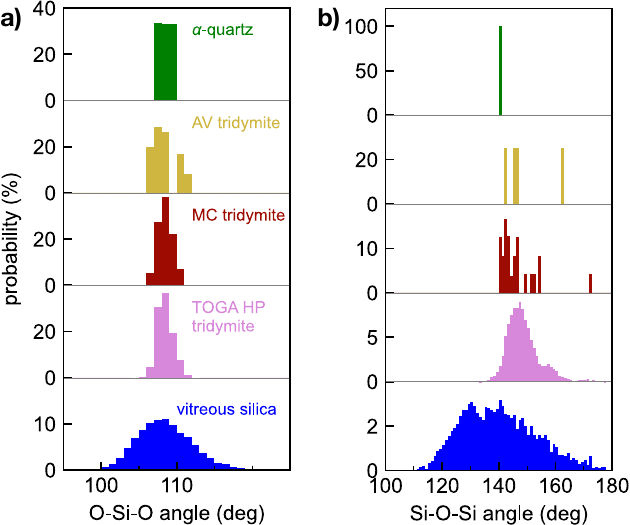}
  \caption{\textbf{Geometric disorder in SiO$_2$ polymorphs.} The probability distribution of having a certain O-Si-O  (Si-O-Si) angle is shown in panel \textbf{a} (\textbf{b}). Broader distributions indicate stronger geometric disorder, which is negatively correlated with the conductivity (Fig.~\ref{fig:k_main_text}).}
  \label{fig:geom_disorder_appendix}
\end{figure}
Fig.~\ref{fig:geom_disorder_appendix} shows the details of the bond geometry of different SiO$_2$ polymorphs. We highlight how, in addition to the dihedral (Si-O-Si-O) angle distribution already shown Fig.~\ref{fig:k_main_text}, also the distribution of the Si-O-Si angle becomes broader as the conductivity decreases. The O-Si-O distribution shows instead smaller variations across different polymorphs, in agreement with the similarity in the shape of the SiO$_4$ tetrahedra in all the polymorphs studied (see snapshots of atomistic structures in Fig.~\ref{fig:k_main_text}\textbf{b)}.

\subsection{Specific heat} 
\label{ssub:specific_heat}
\begin{figure}[h]
  \centering
  \includegraphics[width=\WidthFigure]{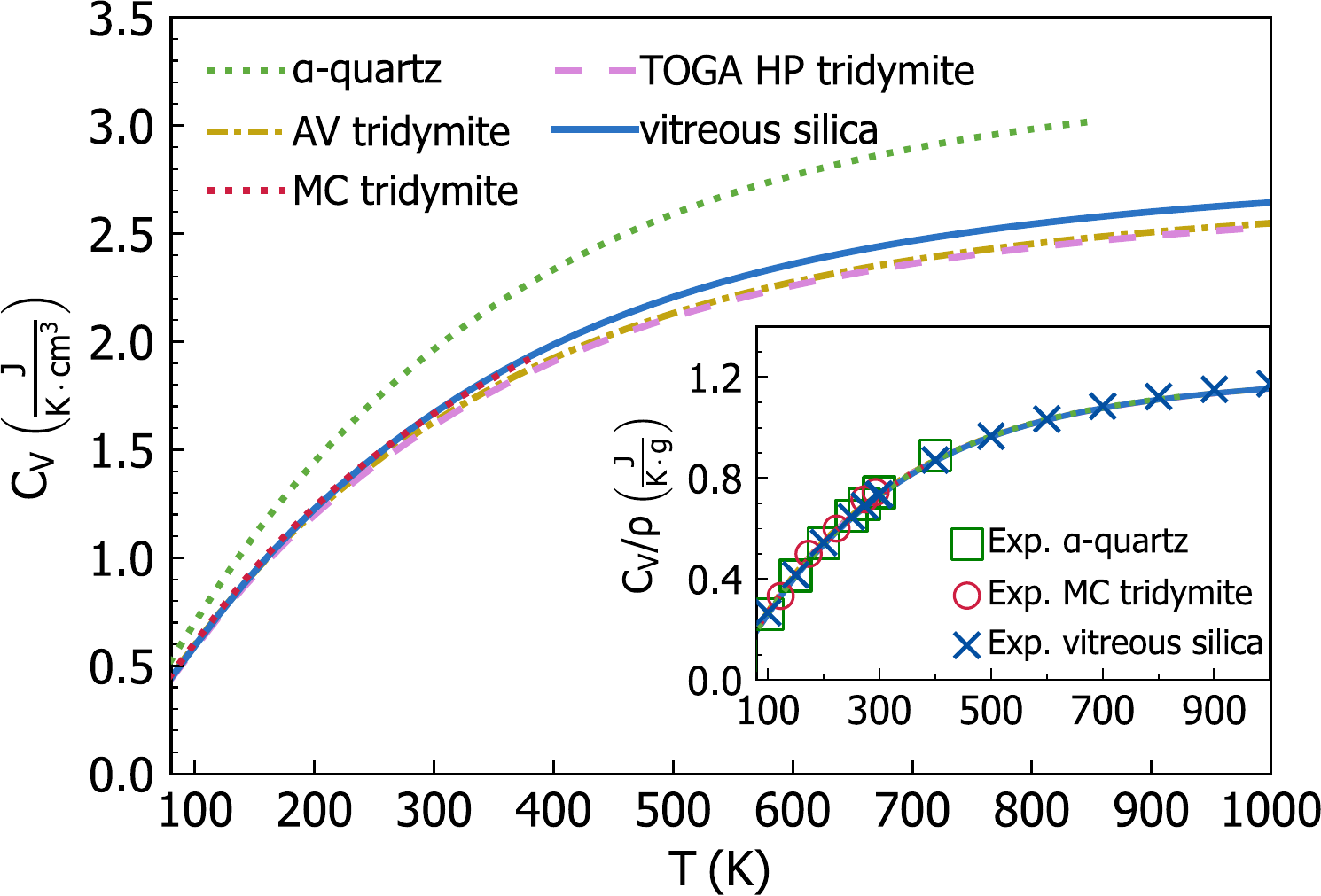}
  \caption{\textbf{Specific heat of silica polymorphs}. lines are the quantum harmonic specific heat at constant volume $C_V$ as a function of temperature; the line for $\alpha$-quartz stops in correspondence of the $\alpha-\beta$ transition at 847 K\cite{richet1982thermodynamic}. The inset shows the specific heat divided by the density $\rho$; scatter points are experimental measurements, for silica and quartz taken from \cite{richet1982thermodynamic}, for MC tridymite taken from \cite{thompson_heat_1979}. }
  \label{fig:spec_heat}
\end{figure}

We show in Fig.~\ref{fig:spec_heat} that the theoretical  quantum
harmonic specific heat at constant volume, $C_V(T){=}\frac{1}{\mathcal{V}N_{\rm c}}\sum_{\bm{q},s} C(\bm{q})_{s}$, is very similar between AV, MC, TOGA HP tridymite and vitreous silica, while $\alpha$-quartz has a larger specific heat. The inset shows that larger specific heat of $\alpha$-quartz originates from its higher density, since the specific heat per unit of mass $C_V(T)/\rho$, where $\rho$ is the density, is practically indistinguishable between all the materials studied, and in close agreement with experimental measurements in vitreous silica and $\alpha$-quartz \cite{richet1982thermodynamic}. 
The good agreement between theoretical predictions for the specific heat at constant volume, and the experiments at constant pressure, shows that the renormalization of vibrational energies due to anharmonicity and temperature \cite{PhysRevB.96.014111,aseginolaza2018phonon,physrevlett.125.085901,physrevb.102.201201,PhysRevB.96.161201,knoop_anharmonicity_2023,PhysRevLett.132.106502} are negligible in these SiO$_2$ polymorphs at the temperatures considered, and as such these effects are not considered in this work.

\subsection{Vibrational frequencies and velocity operators} 
\label{ssub:velocity_operator}
The similarities between the specific heats shown in the previous section motivate us to inspect other possible sources for the differences between macroscopic conductivities shown in Fig.~\ref{fig:k_main_text}. 
Considering that the trends of the macroscopic conductivities do not significantly depend on the Cartesian direction considered, we simplify the description focusing on the average trace of the conductivity tensors. 
In order to gain microscopic insights, we recast the conductivity formula~(\ref{eq:thermal_conductivity_combined}) in terms of physically-insightful frequency-dependent functions:
\begin{equation}
\begin{split}
\kappa=&\int\limits_0^{\omega_{\rm max}}\hspace*{-1mm}d\omega_{\rm a}\hspace*{-1mm}\int\limits_{-\omega_{\rm max}}^{\omega_{\rm max}}\hspace*{-1mm}d\omega_{\rm d}
\Bigg[
\frac{1}{\mathcal{V}N_{\rm c}}{\sum_{\bm{q},s,s'}}
\frac{\omega(\bm{q})_s{+}\omega(\bm{q})_{s'}}{4}\frac{\rVert\tens{v}(\bm{q})_{s,s'}\lVert^2}{3}\\
\times&\!\left[\!\frac{C(\bm{q})_s}{\omega(\bm{q})_s}{+}\frac{C(\bm{q})_{s'\!}}{\omega(\bm{q})_{s'\!}}\!\right]\!\! 
\frac{\frac{1}{2}\big[\Gamma(\bm{q})_s{+}\Gamma(\bm{q})_{s'}\big]}{\big[\omega(\bm{q})_s{-}\omega(\bm{q})_{s'}\big]^2{+}\frac{1}{4}\big[\Gamma(\bm{q})_s{+}\Gamma(\bm{q})_{s'}\big]^2}\\
\times&{\delta}_{\sigma_{\rm a}}\hspace{-1mm}\left(\frac{\omega(\bm{q})_s{+}\omega(\bm{q})_{s'}}{2}{-}\omega_{\rm a}\right) 
{\delta}_{\sigma_{\rm d}}\hspace{-0.5mm}\Big((\omega(\bm{q})_s{-}\omega(\bm{q})_{s'}){-}\omega_{\rm d}\Big)
\Bigg],
\label{eq:thermal_conductivity_rewritten}
\end{split}
\end{equation}
where $\omega_{\rm a}{=}(\omega(\bm{q})_s{+}\omega(\bm{q})_{s'})/2$ and $\omega_{\rm d}{=}\omega(\bm{q})_s{-}\omega(\bm{q})_{s'}$ are the average and difference between the frequencies of vibrational eigenstates coupled by off-diagonal velocity-operator elements, and the distributions $\delta_{\sigma}$ are Gaussian broadenings of the Dirac delta: 
\begin{equation}
  \delta_{\sigma}\left(\Omega-\omega\right)=\frac{1}{\sqrt{2\pi}\sigma} \exp\left[{-\frac{1}{2\sigma^2}\left(\Omega-\omega\right)^2}\right].
\end{equation}
In order to simplify the analysis, we approximate the linewidths using the single-valued function of frequency\cite{simoncelli_thermal_2023,harper_vibrational_2024,Anees_2023}, \textit{i.e.} $\Gamma(\bm{q})_s=\Gamma_{\rm a}[\omega(\bm{q})_s]$; using this approximation allows to recast Eq.~(\ref{eq:thermal_conductivity_rewritten}) as follows:
\begin{equation}
\begin{split}
&\kappa=\int\limits_0^{\omega_{\rm max}}d\omega_{\rm a}\int\limits_{-\omega_{\rm max}}^{\omega_{\rm max}}d\omega_{\rm d}
\mathcal{G}(\omega_{\rm a},\omega_{\rm d})\mathcal{C}(\omega_{\rm a},\omega_{\rm d})
\left<|V^{\rm avg}_{\omega_{\rm a},\omega_{\rm d}}|^2\right>\\
&\hspace*{15mm}\times
\frac{\frac{1}{2}\big(\Gamma_{\rm a}[\omega_{\rm a}{+}\frac{\omega_{\rm d}}{2}]{+}\Gamma_{\rm a}[\omega_{\rm a}{-}\frac{\omega_{\rm d}}{2}]\big)}{\omega_{\rm d}^2{+}\frac{1}{4}\big(\Gamma_{\rm a}[\omega_{\rm a}{+}\frac{\omega_{\rm d}}{2}]{+}\Gamma_{\rm a}[\omega_{\rm a}{-}\frac{\omega_{\rm d}}{2}]\big)^2}
\label{eq:thermal_conductivity_rewritten2}
  \raisetag{8mm}
\end{split}
\end{equation}
where $\mathcal{G}(\omega_{\rm a},\omega_{\rm d})$ is a density of states
\begin{equation}
\begin{split}
  \mathcal{G}(\omega_{\rm a},\omega_{\rm d})
  =\frac{1}{N_{\rm at}}\frac{1}{\mathcal{V}N_{\rm c}}&{\sum_{\bm{q},s,s'}}{\delta}_{\sigma_{\rm a}}\left(\frac{\omega(\bm{q})_s+\omega(\bm{q})_{s'}}{2}-\omega_{\rm a}\right) \\
 \times&{\delta}_{\sigma_{\rm d}}\big((\omega(\bm{q})_{s}{-}\omega(\bm{q})_{s'})-\omega_{\rm d}\big)\;,
  \label{eq:2freq_vDOS}
  \raisetag{5mm}
\end{split}
\end{equation}
$\mathcal{C}(\omega_{\rm a},\omega_{\rm d})$ is a specific heat 
\begin{equation}
  \mathcal{C}(\omega_{\rm a},\omega_{\rm d})=\frac{\omega_{\rm a}}{2}\left(\frac{C(\omega_{\rm a}+\frac{\omega_{\rm d}}{2})}{\omega_{\rm a}+\frac{\omega_{\rm d}}{2}}+\frac{C(\omega_{\rm a}-\frac{\omega_{\rm d}}{2})}{\omega_{\rm a}-\frac{\omega_{\rm d}}{2}}\right),\;
  \label{eq:2freq_spec_heat}
\end{equation}
and $\left<|\tenscomp{v}^{\rm avg}_{\omega_{\rm a},\omega_{\rm d}}|^2\right>$ is the average square modulus of the velocity operator
\begin{equation}
\begin{split}
  &\big<|\tenscomp{v}^{\rm avg}_{\omega_{\rm a}\omega_{\rm d}}|^2\big>=\bigg[[\mathcal{G}(\omega_{\rm a},\omega_{\rm d})]^{-1}\frac{1}{\mathcal{V}N_{\rm c}}{\sum_{\bm{q},s,s'}} \frac{\rVert\tens{v}(\bm{q})_{s,s'}\rVert^2}{3}\\
  &\hspace*{1cm}{\times} 
 \delta\left(\omega_{\rm d}{-}(\omega(\bm{q})_s{-}\omega(\bm{q})_{s'})\right)\delta\Big(\omega_{\rm a}{-}\frac{\omega(\bm{q})_s{+}\omega(\bm{q})_{s'}}{2} \Big)\bigg]\;.
 \raisetag{20mm}
  \label{eq:v_operator_omega_a_omega_d}
\end{split}
\end{equation}

\begin{figure}[ht!]
  \includegraphics[width=\WidthFigure]{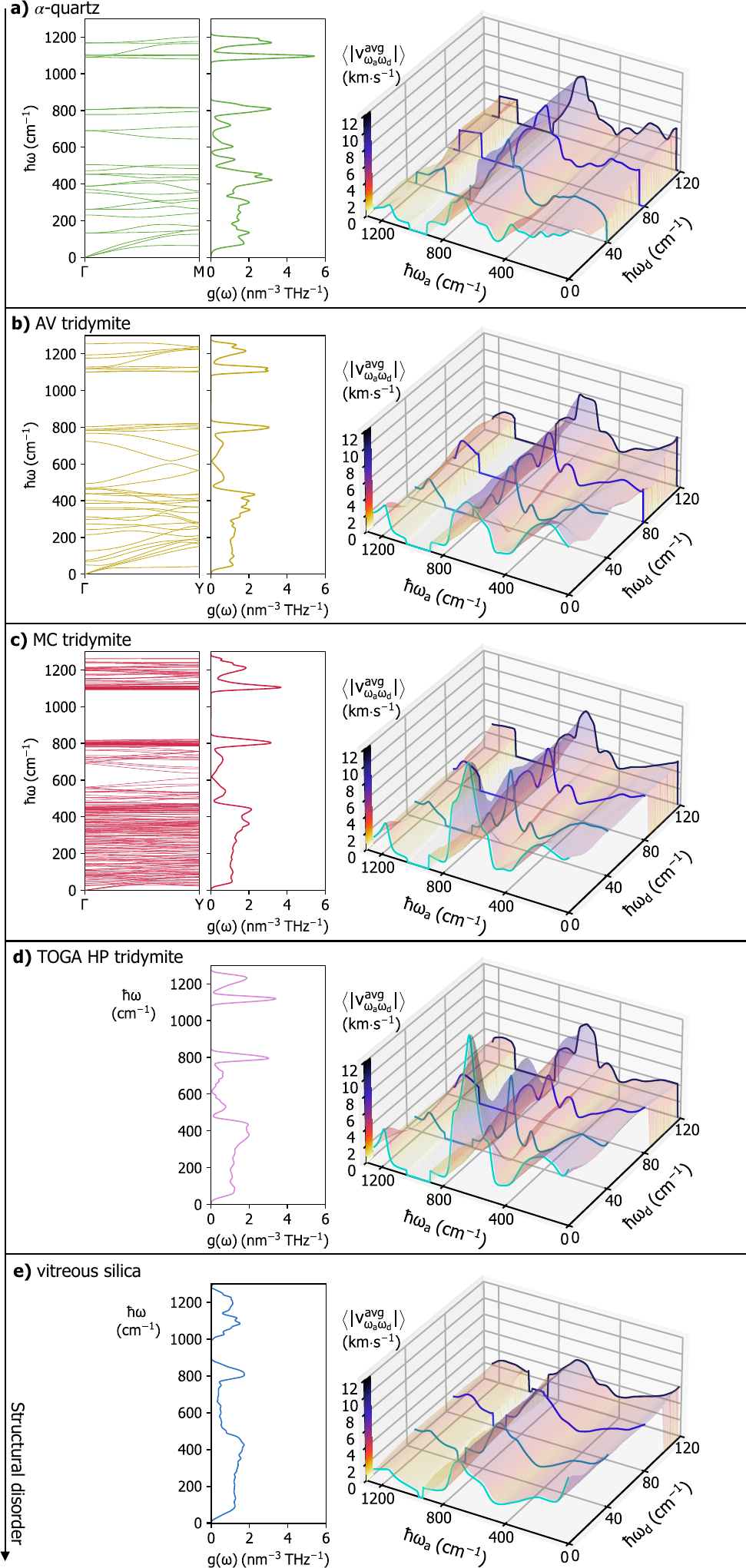}
  \caption{\textbf{Vibrational frequencies and velocity operators of SiO$_2$ polymorphs with various degree of disorder.} Structural disorder increases from top to bottom. The left column shows the vibrational properties (the phonon dispersion relation is shown for the crystalline phases only, the  vibrational density of states is always shown). The right column shows the velocity operator as function of  energy difference $\hbar\omega_{\rm d}{=}\hbar(\omega(\bm{q})_{s}{-}\omega(\bm{q})_{s'})$ and average $\hbar\omega_{\rm a}{=}\hbar\frac{\omega(\bm{q})_{s}{+}\omega(\bm{q})_{s'}}{2}$ of the two modes couped $(\bm{q})_s$ and $(\bm{q})_{s'}$, computed using Eq.~(\ref{eq:v_operator_omega_a_omega_d}).\vspace*{-5mm}}
  \label{fig:velop}
\end{figure}

Overall, the analysis above shows that for polymorphs having practically indistinguishable specific heat (Fig.~\ref{fig:spec_heat}) and similar anharmonicity~(Fig.~\ref{fig:lifetime}, see also Ref.~\cite{mizokami_lattice_2018} for a discussion on the similarity of anharmonicity across different silica polymorphs), the conductivity differences are mainly determined by differences in the velocity operators elements. For ordered polymorphs such as AV tridymite, only the terms with $\omega_{\rm d}=0$ are relevant and have a strength inversely proportional to the linewidth. In contrast, for disordered polymorphs also elements with $\omega_{\rm d}\neq 0$ contribute to conduction, and as shown by Eq.~(\ref{eq:thermal_conductivity_rewritten2}), the conductivity is a weighted average of velocity-operator elements. We will show later in Fig.~\ref{fig:eff_anh} that in the most disordered polymorphs TOGA HP tridymite and vitreous silica, 
anharmonicity has weak (TOGA HP) or negligible (vitreous silica)  effects and 
the conductivity is mainly determined by disorder. 

Fig.~\ref{fig:velop} shows the harmonic vibrational properties and velocity operators $\big<|\tenscomp{v}^{\rm avg}_{\omega_{\rm a}\omega_{\rm d}}|\big>=\sqrt{\big<|\tenscomp{v}^{\rm avg}_{\omega_{\rm a}\omega_{\rm d}}|^2\big>}$ of the polymorphs studied (the plot was computed using Eq.~(\ref{eq:v_operator_omega_a_omega_d}) with $\sigma_{\rm a}=21$ cm$^{-1}$ and $\sigma_{\rm d}=7$ cm$^{-1}$ for all polymorphs). The vibrational density of states is computed using the usual expression,  $g(\omega){=}({\mathcal{V}N_{\rm c}})^{-1}\sum_{\bm{q},s}\delta(\omega{-}\omega(\bm{q})_s)$, where $\mathcal{V}$ is the primitive cell volume, $N_c$ is the number of $\bm{q}$ points appearing in the summation, and $\hbar\omega(\bm{q})_s$ the energy of the vibrational mode with wavevector $\bm{q}$ and mode $s$.
We highlight how increasing geometric disorder at fixed bond topology (AV $\to$ MC $\to$ TOGA HP) yields an increase in off-diagonal velocity operator elements, explaining the increase in the corresponding coherences conductivities shown in Fig.~\ref{fig:lifetime}. We also note that relaxing the constraint of preserving crystalline bond topology in a geometrically disordered material  yields a reduction in off-diagonal velocity-operator elements, explaining why TOGA HP tridymite displays a conductivity larger than vitreous silica (Fig.~\ref{fig:k_main_text}).

\subsection{Effects of anharmonicity on the conductivity} 
\label{sub:thermal_conductivity_of_tridymite_polymorphs}
\begin{figure}[b]
  \centering
  \includegraphics[width=\WidthFigure]{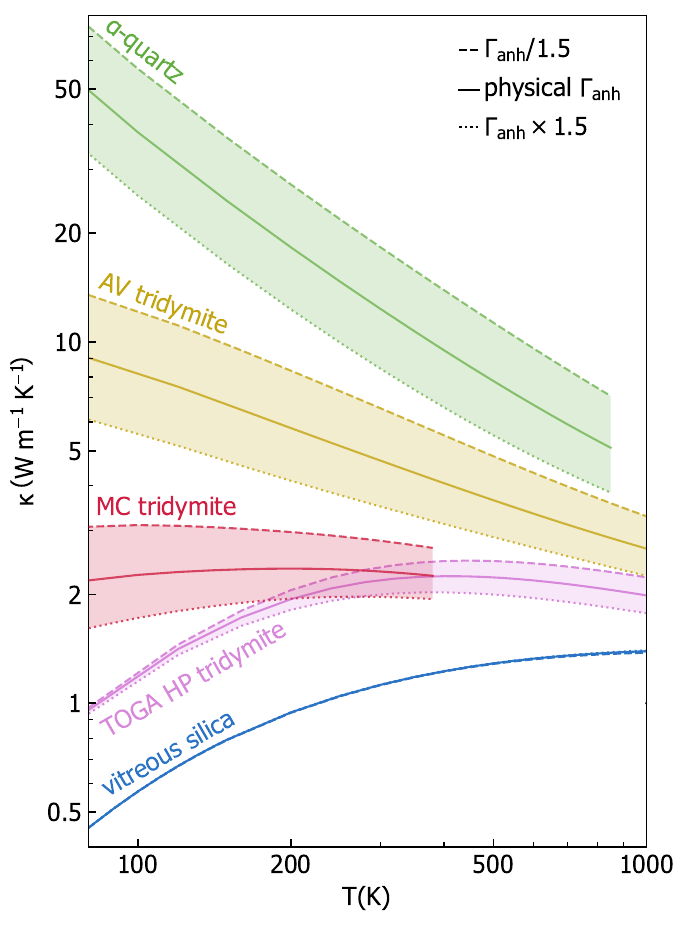}\vspace*{-5mm}
\caption {\textbf{Effect of anharmonicity on the thermal conductivity of silica polymorphs.} 
Solid lines are the thermal conductivities computed using the physical values of the anharmonic linewidths, also shown in Fig.~\ref{fig:k_main_text}.
The dashed (dotted) lines are obtained using anharmonic linewidth artificially divided (multiplied) by 1.5; the shaded area between these lines shows how sensitive is the total thermal conductivity to the anharmonic linewidths. In ordered materials such a $\alpha$-quartz, the conductivity is determined by the anharmonic linewidths, while in disordered materials such as silica glass the total thermal conductivity is determined by structural disorder and practically independent from the linewidths.
} \label{fig:eff_anh}
\end{figure}

In this section we investigate how sensitive the macroscopic thermal conductivity~(\ref{eq:thermal_conductivity_combined}) is to anharmonicity, inspecting how much it varies when the linewidths are artificially enlarged or reduced by a factor $1.5$.
Fig.~\ref{fig:eff_anh} shows that simple crystals such as $\alpha$-quartz and AV tridymite, in which heat transport is mainly due to particle-like mechanisms (see Ref.~\cite{simoncelli_thermal_2023} and Fig.~\ref{fig:lifetime}), the artificial decrease (increase) of the linewidths directly translates into an increase (decrease) of the conductivity. This behavior follows trivially from the expression for the particle-like conductivity discussed in the main text, $\kappa_P^{\alpha\beta}{=}\frac{1}{\mathcal{V} N_C }\sum_{\bm{q}s}C[\omega(\bm{q})_{s}]{\tenscomp{v}^{\alpha}(\bm{q})_{s,s}}{\tenscomp{v}^{\beta}(\bm{q})_{s,s}}\Gamma(\bm{q})_s$. 

In MC tridymite the conductivity variations upon artificial rescaling of anharmonic linewidths follow the same qualitative behavior, albeit these have a magnitude that is overall reduced compared to those observed in $\alpha$-quartz and AV tridymite. Such a reduced sensitivity to anharmonicity in  MC tridymite originates from: (i) having a conductivity determined by both particle-like and wave-like mechanisms; (ii) having strong effect of artificial variations of anharmonicity on the particle-like conductivity, and weak effect on the wave-like conductivity. 
At high temperature, wave-like conduction mechanisms give the strongest contribution to the conductivity of MC tridymite; this, in addition to the points above, implies that artificial variations of anharmonicity have a weaker impact on the conductivity of MC tridymite as temperature increases. 

Increasing geometric disorder at fixed bond topology (MC$\to$TOGA HP) particle-like transport is suppressed (Fig.~\ref{fig:lifetime}) and wave-like transport enhanced (Fig.~\ref{fig:velop}).
As discussed before, when wave-like transport dominates, the conductivity can be resolved in terms of weighted average of non-degenerate velocity operator elements. As shown by Eq.~(\ref{eq:v_operator_omega_a_omega_d}), increasing the anharmonic linewidths corresponds to giving more weight to elements with larger energy difference $\hbar\omega_d$.
Therefore, the variation of the velocity operator with with respect to $\hbar\omega_d$  determines how the conductivity changes upon varying the linewidths: matrix elements increasing (decreasing) with $\hbar\omega_d$ imply a conductivity increasing (decreasing) upon augmenting the linewidths. Fig.~\ref{fig:velop}\textbf{d)} shows that the velocity operator of TOGA HP tridymite is approximatively constant with respect to $\omega_{\rm d}$ for $\omega_{\rm a}\lesssim 400$ cm$^{-1}$, and it decreases significantly with  $\hbar\omega_d$ for $400\lesssim\omega_{\rm a}\lesssim 800$ cm$^{-1}$. 
At low temperature, where only the low-frequency vibrational modes are significantly populated ($\bar{\tenscomp{N}}(\omega_{\rm a}){\gg}0 {\iff} \hbar\omega_{\rm a}{\ll} k_B T$), artificially rescaling the anharmonic linewidths produces negligible effects, since the velocity operator is approximatively constant with respect to $\omega_{\rm d}$ for the low-energy modes that are non-negligibly populated.
As temperature increases, vibrational modes in the energy range $400{\lesssim}\omega_{\rm a}{\lesssim} 800$ cm$^{-1}$ become populated, and in this region increasing (decreasing) linewidths corresponds to giving more (less) weight to smaller velocity-operator elements; therefore, artificially increasing (decreasing) the linewidths yields a reduction (augmentation) in the conductivity of TOGA HP tridymite at high temperature. 

We conclude by highlighting how the trend of the velocity operator as a function of $\omega_{\rm d}$, and thus the effect of varying the anharmonic linewidths, depends on the  details of the atomistic structure. In particular, Fig.~\ref{fig:velop}\textbf{d)} shows that in vitreous silica  the velocity operator is practically independent from $\omega_{\rm d}$ for all values of $\omega_{\rm a}$ \cite{simoncelli_thermal_2023}, implying that in this case the conductivity is practically unaffected by the artificial rescaling of the linewidths.

\subsection{Compensation of particle-like and wave-like thermal conductivity in MC tridymite} 
\begin{figure}[b]
  \centering
  \includegraphics[width=\WidthFigure]{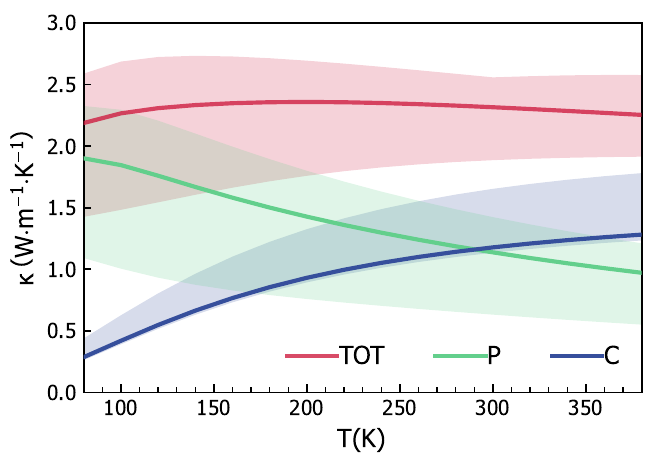}\\[-5mm]
  \caption{\textbf{Compensation of propagation and tunneling conductivity in MC tridymite.} 
Upon increasing temperature, the increase of the wave-like-tunneling conductivity (blue, $\kappa_C$) is as fast as the decrease of the particle-like-propagation conductivity (green, $\kappa_P$), yielding a nearly temperature-invariant total conductivity (red, $\kappa_{TOT}=\kappa_P+\kappa_C$). The shaded area shows the maximum and minimum eigenvalues of the corresponding conductivity tensor. }
  \label{fig:balancing}
\end{figure}
In Fig.~\ref{fig:balancing} we decompose the total thermal conductivity of MC tridymite reported in Fig.~\ref{fig:k_main_text} in terms of particle-like and wave-like contributions.
We highlight how such decomposition is well defined in MC tridymite\cite{simoncelli_wigner_2022}, since it is a complex crystal with translation-invariant symmetry, well-defined primitive cell\cite{Dollase76} and phonon dispersion\cite{hinuma_band_2017}. This is a fundamental difference compared to other, common types of disorder that break translational symmetry (e.g. compositionally disordered alloys~\cite{cahill_lower_1992,Anees_2023,thebaud_breaking_2023} or amorphous solids\cite{harper_vibrational_2024,simoncelli_thermal_2023,fiorentino_hydrodynamic_2023}), which do not allow to rigorously define a phonon band structure. In the context of Ref.~\cite{simoncelli_wigner_2022}, these systems with broken translational symmetry  and no perfectly degenerate vibrational modes feature wave-like heat transport only.

\subsection{Strength of propagation \& tunneling transport}

Here we discuss the details underlying the investigation of the relative strength of microscopic particle-like propagation and wave-like tunneling transport mechanisms reported in Fig.~\ref{fig:lifetime}.
As discussed in Ref.~\cite{simoncelli_wigner_2022}, it is possible to resolve how much each vibrational mode $(\bm{q})_s$ contributes to the particle-like and wave-like conductivities. Such information is contained in the mode-resolved contributions to the particle-like conductivity tensor, 
\begin{equation}
{\mathcal{K}}^{\alpha\beta}_{P}(\bm{q})_{s} = C(\bm{q})_s \tenscomp{v}^{\alpha}(\bm{q})_{s,s}\tenscomp{v}^{\beta}(\bm{q})_{s,s} \left [ \Gamma(\bm{q})_s \right ]^{-1}, \label{eq:kp-av}
\end{equation}
and in the mode-resolved contributions to the wave-like conductivity tensor,  
\begin{equation}
  \begin{split}
&{\mathcal{K}}^{\alpha\beta}_C(\bm{q})_s{=} \!\!
\sum_{s'\neq s}\!\!
  \frac{C(\bm{q})_s}{C(\bm{q})_{s}{+}C(\bm{q})_{s'\!}\!}
\frac{\omega(\bm{q})_{s}{+}\omega(\bm{q})_{s'}\!}{2}\!\!
\left[\!\frac{C(\bm{q})_s}{\omega(\bm{q})_s}{+}\frac{C(\bm{q})_{s'\!}}{\omega(\bm{q})_{s'\!\!}}\!\right]\!\\
&{\times}
\tenscomp{v}^{\alpha}(\bm{q})_{s,s'\!}\tenscomp{v}^{\beta}(\bm{q})_{s,s'\!} \frac{\tfrac{1}{2}\big[\Gamma(\bm{q})_{s}{+}\Gamma(\bm{q})_{s'}\big]}{[\omega(\bm{q})_{s'}{-}\omega(\bm{q})_{s}]^2{+}\tfrac{1}{4}[\Gamma(\bm{q})_{s}{+}\Gamma(\bm{q})_{s'}]^2}.
  \end{split}
\label{eq:coherence_density}
  \end{equation}

Ref.~\cite{simoncelli_wigner_2022} demonstrated that the relative strength of these contributions scales as the ratio between the average energy-level spacing  ($\hbar\Delta\omega_{\rm avg} {=} \frac{\hbar\omega_{\rm max}}{3N_{at}}$, where  $\hbar\omega_{\rm max}$ is the maximum vibrational energy, $N_{at}$ is the number of atoms in the primitive cell, and $3N_{at}$ the number of energy levels) and the linewidth
$\Gamma(\bm{q})_{s}{=}[\tau(\bm{q})_{s}]^{-1}$ (here $\tau(\bm{q})_{s}$ is the phonon lifetime). In formulas,
\begin{equation}\label{eq1}
  \frac{{\mathcal{K}}^{\alpha\beta}_{C}(\bm{q})_{s}}{{\mathcal{K}}^{\alpha\beta}_{P}(\bm{q})_{s}} \simeq \frac{\Gamma(\bm{q})_{s}}{\Delta\omega_{\rm avg}} = \frac{[\Delta\omega_{\rm avg}]^{-1}}{\tau(\bm{q})_{s}}.
\end{equation}
Eq.~(\ref{eq1}) predicts that vibrations having a lifetime $\tau(\bm{q})_s$ equal to the inverse interband spacing $[\Delta\omega_{\rm avg}]^{-1}$ (also referred to as `Wigner limit in time'\cite{simoncelli_wigner_2022,caldarelli2022}) contribute simultaneously and with equal strength to both particle-like and wave-like conduction mechanism.
In contrast, vibrations with a lifetime much longer (shorter) than the Wigner limit in time contribute predominantly to the particle-like (wave-like) conductivity. 
Finally, Eq.~(\ref{eq1}) predicts  the transition between these two limits to be non-sharp and centered at the Wigner limit in time. 
These analytical expectations are verified numerically in Fig.~\ref{fig:lifetime}. 
Each scatter point in Fig.~\ref{fig:lifetime} provides information on the conduction mechanisms through which the corresponding vibrational mode participates to heat transport, as well as on how much the microscopic phonon mode contributes to the macroscopic conductivity.
The first information on the type of conduction mechanisms is encoded in the color of the scatter point, determined according to the value of the parameter
 \begin{equation}\label{eq3}
   c=\frac{{\mathcal{K}}^{\alpha\beta}_{P}(\bm{q})_{s}-{\mathcal{K}}^{\alpha\beta}_{C}(\bm{q})_{s}}{{\mathcal{K}}^{\alpha\beta}_{P}(\bm{q})_{s}+{\mathcal{K}}^{\alpha\beta}_{C}(\bm{q})_{s}}.
\end{equation} 
Eq.~(\ref{eq3}) implies that $c{=}{+}1$ when the vibration $(\bm{q})_s$ predominantly contributes to particle-like conduction (corresponding to green),  $c{=}{-}1$ when instead the vibration contributes mainly to wave-like conduction (blue), and finally $c{=}0$ when the vibration contributes equally to particle-like and wave-like conduction (red).
We have employed a linear color scale interpolating blue, red, and green to resolve all the possible hybrid cases. The second information, on the magnitude of the transport mechanisms, is represented by the area of each scatter point, which is proportional to the contribution of such vibration to the total thermal conductivity. 
Finally, we highlight that all vibrations here have a lifetime longer than their reciprocal frequency ($\tau(\bm{q})_s>[\omega(\bm{q})_s]^{-1}$), \textit{i.e.} they are all above the dashed-purple areas. Therefore, Landau's quasiparticle picture  \cite{landau1980statistical} for vibrational excitations holds for the silica polymorphs investigated here, and consequently the Wigner formulation can be applied \cite{simoncelli_wigner_2022,caldarelli2022}.

\section{Computational details} 
\label{sec:computational_details}

The GAP potential for silica polymorphs is taken from Ref.~\cite{erhard2022machine}.
Structure relaxation, interatomic forces, and stress tensor are computed using \texttt{LAMMPS}\cite{brown2011implementing}, using the \texttt{quip}\cite{Csanyi2007-py,GAP_potential,Kermode2020-wu} interface to call the GAP potential routines. 
Cell parameters and atomic positions are relaxed using a threshold of $25$ eV/Angstrom for the atomic forces (\textit{i.e.} a structure is considered relaxed if all the Cartesian components of the forces acting on atoms are smaller than this threshold), and a threshold of 0.1 kBar for pressure. 
The harmonic dynamical matrices (which yield the vibrational frequencies and velocity operators) are computed using \texttt{Phonopy}\cite{togo_implementation_2023}.
Third-order force constants are computed using \texttt{ShengBTE} \cite{li2014shengbte}, using a cutoff equal to the $7^{\rm th}$ nearest-neighbor for all the polymorphs; the resulting force constants are then converted in \texttt{Phono3py}\cite{togo_implementation_2023,phono3py} format using \texttt{hiphive}\cite{hiphive}.
Linewidths are computed accounting for third-order anharmonicity \cite{paulatto2013anharmonic,fugallo2013ab} and isotopic disorder at natural abundance \cite{tamura1983isotope}. The computational details for every polymorph studied are reported below.

\subsection{$\alpha$-quartz} 
\label{sub:subsection_name}

The crystal structure is taken from Ref. \cite{quartz_structure} (Crystallographic Open Database id 1526860). 
After relaxing the atomic positions and the lattice parameters, the density is 2.6801 g/cm$^3$, in good agreement with the experimental value\cite{2014crc} 2.65 g/cm$^3$.
Harmonic force constants are computed on a $6{\times}6{\times}5$ supercell in real space, anharmonic third-order force constants are computed using a $3{\times}3{\times}2$ supercell. The thermal conductivity is computed on a  $19{\times}19{\times}15$ $\bm{q}$ mesh and using a Gaussian smearing of $0.4$ cm$^{-1}$ to numerically evaluate the Dirac delta appearing in the linewidth expression\cite{phono3py}. Results are practically indistinguishable from those obtained from first principles in Ref.~\cite{simoncelli_thermal_2023}.
\\

\subsection{AV tridymite} 
\label{sub:AV_T}
The crystal structure is taken from Ref.~\cite{graetsch_29si_nodate}. 
After relaxing the atomic positions and the lattice parameters, the density is 2.2066 g/cm$^3$.
Harmonic force constants are computed on a $5{\times}5{\times}3$ supercell, anharmonic third-order force constants are computed using a $3{\times}3{\times}2$ supercell. The thermal conductivity is computed on a  $9{\times}9{\times}7$ $\bm{q}$ mesh and using a Gaussian smearing of $0.4$ cm$^{-1}$ to numerically evaluate the Dirac delta appearing in the linewidth expression\cite{phono3py}. We have verified that increasing the $\bm{q}$ mesh size to $19{\times}19{\times}11$ produces results compatible (within 4.5\%) with those obtained using the  $9{\times}9{\times}7$ $\bm{q}$ mesh. \\

\subsection{Monoclinic MC tridymite} 
The crystal structure is taken from Ref.~\cite{Dollase76}.
After relaxing the atomic positions and the lattice parameters, the density is 2.2676 g/cm$^3$.
Harmonic force constants are computed on a $3{\times}3{\times}2$ supercell in real space, anharmonic third-order force constants are computed using a $2{\times}2{\times}1$ supercell. The thermal conductivity is computed on a  $13{\times}13{\times}3$ $\bm{q}$ mesh and using a Gaussian smearing of $0.4$ cm$^{-1}$ to numerically evaluate the Dirac delta appearing in the linewidth expression\cite{phono3py}.
The solid red line shown in Fig.~\ref{fig:k_main_text} is computed along the direction of the vector $\bm{m} = 0.65 \bm{a} - 4.7 \bm{b} + 0.715 \bm{c}$, where $\bm{a}$, $\bm{b}$, $\bm{c}$ are the direct lattice vectors of the structure discussed by Dollase \textit{et al.}\cite{Dollase76}. Specifically, the effective conductivity in direction $\bm{\hat m}=\frac{\bm{m}}{\sqrt{\bm{m}\cdot \bm{m}}}$ was computed considering a temperature gradient in direction $\bm{\hat m}$, computing the heat flux along such direction, projecting such heat flux along the temperature-gradient direction, and then dividing by the magnitude of the temperature gradient, yielding
  $\kappa_{\bm{\hat m}}={\sum_{\alpha,\beta} \hat{m}^\alpha \kappa^{\alpha\beta}\hat {m}^\beta}$. The shaded red area in Fig.~\ref{fig:k_main_text} shows the difference between the minimum and the maximum eigenvalue of the thermal conductivity tensor at a given temperature, and is indicative of how much the thermal conductivity depends from the relative orientation between crystallographic axes and direction of the temperature gradient. \\

\subsection{TOGA HP tridymite} 
{The structure is generated starting from a $9{\times}9{\times}7$ supercell of AV tridymite, which we showed in Sec.~\ref{sub:AV_T} to be sufficiently large to yield computational convergence in the thermal conductivity calculation of crystalline AV tridymite.
Geometric disorder is created via a molecular dynamics simulation at 1000 K. Specifically, the structure is heated from 0 to 1000 K (NPT) in 2 ps, and then equilibrated at 1000 K for 1 ns (NPT). 
No constraint are imposed on the shape and size of the cell during the simulation. The timestep is 0.001 ps, and no bonds are broken during the simulation. After the simulation, the structure of TOGA HP tridymite is obtained by relaxing the the atomic positions and the lattice parameters of the final snapshot at 1000 K. Such relaxation is performed using a threshold of $25$ eV/Angstrom for the atomic forces, and of 0.1 kBar for pressure. This procedure yields a density equal to 2.1920 g/cm$^3$. Moreover,  we show in Tab.~\ref{tab:struct_HP} that the average crystal structure of TOGA HP tridymite (obtained dividing by 9,9, and 7 the supercell lattice vectors) is compatible with the hexagonal crystal symmetry of the idealized HP tridymite structure\cite{graetsch_modulated_2009}, confirming that our TOGA HP tridymite is a realistic representation of the high-temperature HP tridymite phase.

\begin{table}[h]
  \caption{Cell vectors and angles of AV tridymite, and of the average cell of TOGA HP tridymite. Both cells contain 12 atoms. We highlight how the average cell of TOGA HP tridymite is very close to the hexagonal crystal symmetry of the idealized 12-atom primitive cell of HP tridymite\cite{graetsch_modulated_2009}.}
  \label{tab:struct_HP}
  \centering

  \begin{tabular}{l|l}
  \hline

    \hline
  \textbf{lattice vectors} & \textbf{angles}  \\
  \hline
\multicolumn{2}{c}{\textbf{AV tridymite}}\\
  \hline
$|\bm{a}|=$ 5.013126   & $\alpha$= 90.761 \\
$|\bm{b}|=$5.013126     & $\beta$=  90.761  \\
$|\bm{c}|=$8.278431 & $\gamma$=   119.586 \\
  \hline
\multicolumn{2}{c}{\textbf{Average cell of TOGA HP tridymite}}
 \\
  \hline
$|\bm{a}|=$5.037416   & $\alpha$= 90.031 \\
$|\bm{b}|=$5.032923    & $\beta$=  90.316  \\
$|\bm{c}|=$8.298937 & $\gamma$=  120.081 \\
  \hline

  \hline
  \end{tabular}
\end{table}

The thermal conductivity is computed at $\bm{q}=\bm{0}$ only, corresponding to a system containing 6804 atoms, a size equal to that employed for the computationally converged thermal conductivity calculation  of crystalline AV tridymite.
The thermal conductivity of TOGA silica presented in Fig.~\ref{fig:k_main_text} was evaluated considering the anharmonic linewidths of AV tridymite computed over a $9{\times}9{\times}7$ $\bm{q}$-mesh. The linewidths of AV tridymite were assigned to modes in TOGA silica using a frequency-similarity protocol, \textit{i.e.} for every vibrational mode of TOGA HP tridymite $\omega^{\rm TOGA-HP}(\bm{0})_s$, the linewidth was determined by first finding the mode of AV tridymite $\omega^{\rm AV}(\bm{q})_s$  belonging to a $9{\times}9{\times}7$ $\bm{q}$-mesh and having the closest frequency to $\omega^{\rm TOGA-HP}(\bm{0})_s$; thus assigning the linewidth of the mode $\Gamma^{\rm AV}(\bm{q})_s)$ to the one with closest frequency in TOGA HP tridymite. The accuracy of such an approximation is validated in Fig.~\ref{fig:eff_anh}, where we show that artificially enlarging or reducing the linewidths factor 1.5 produces unimportant changes on the conductivity of TOGA HP tridymite. This indicates that the conductivity of TOGA HP tridymite is mainly determined by geometric disorder, and thus the approximated treatment of anharmonicity employed in TOGA HP tridymite is sufficiently accurate for the scope of the present study.
Finally, to confirm that our TOGA HP tridymite model is a realistic representation of HP tridymite, we show in Fig.~\ref{fig:hexagonal_sym} that the thermal conductivity tensor of TOGA HP tridymite practically respects the hexagonal average crystal symmetries (which were expected from the structural properties shown in Tab.~\ref{tab:struct_HP}).
\begin{figure}[h]
  \centering
  \includegraphics[width=\WidthFigure]{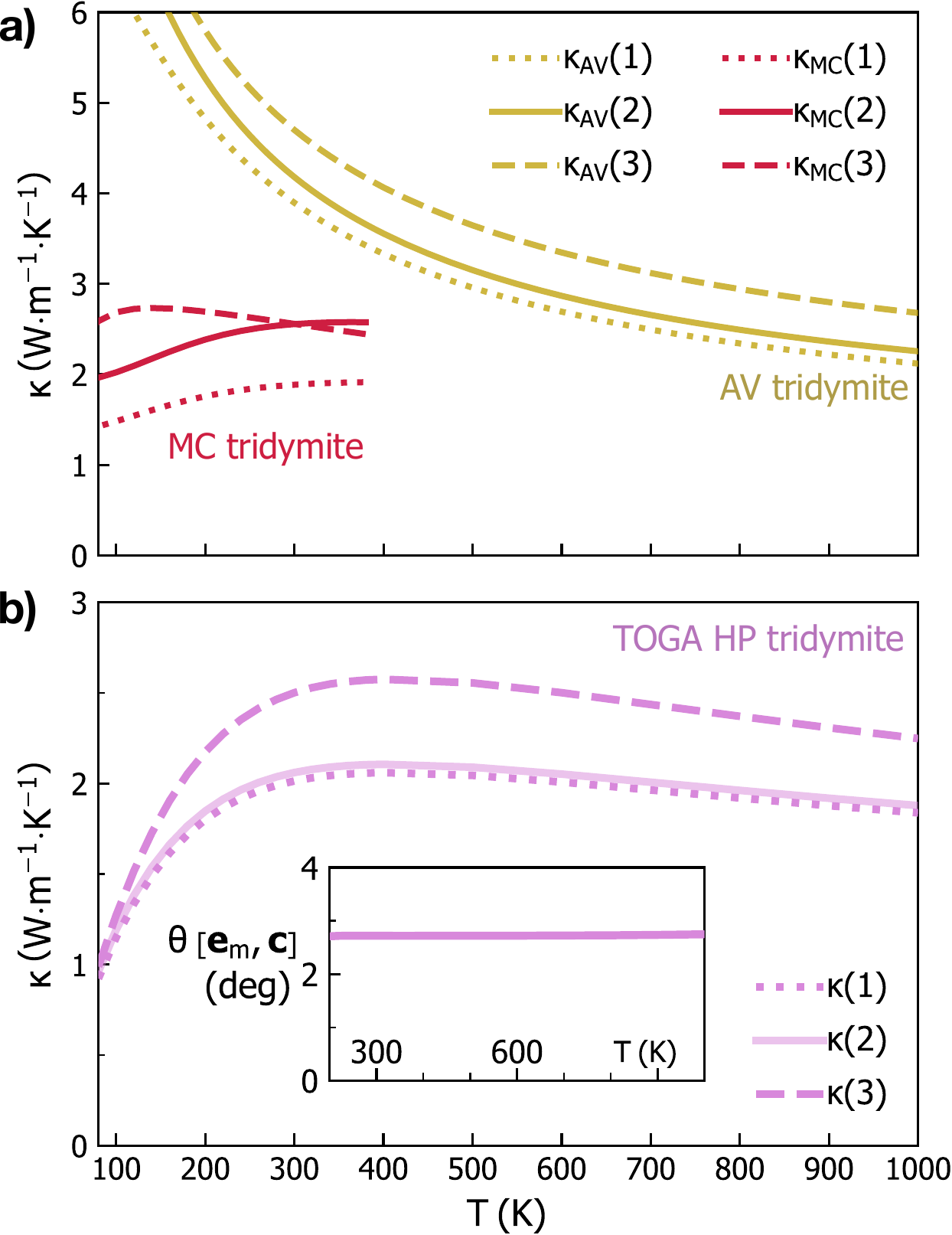}
  \caption{\textbf{Average hexagonal symmetry of the conductivity tensor in TOGA HP tridymite.} 
  Dashed, solid, and dotted line are the 3 eigenvalues of the conductivity tensor: panel \textbf{a)} shows AV tridymite (yellow) and MC tridymite (red); panel \textbf{b)} shows TOGA HP tridymite (pink). The inset in \textbf{b)} shows the angle between the eigenvector corresponding to the largest eigenvalue and the longest ($\bm{c}$) lattice vector. 
  }
  \label{fig:hexagonal_sym}
\end{figure}

The experimental data on the thermal conductivity of TOGA HP tridymite (purple empty diamonds in Fig.~\ref{fig:k_main_text}) are determined from Ref.~\cite{Pierce_1936}. 
Specifically, after digitalizing the curve `A' in Fig.~6 of such reference (temperature as a function of distance from the hot end  $d$ for a 9.52-inch uninsulated brick), we computed the temperature gradient $\nabla T(d)$ as a function of  $d$. 
Ref.~\cite{Pierce_1936} states that the heat flux flowing through the refractory brick is $|Q|=$ 3200 BTU/(hour$\cdot$ foot$^2$) (corresponding to 10094.69 W/mK), and also reports the porosity of the sample $\phi$ as a function of $d$. 
 We rely on such data to obtain the conductivity as a function of $d$, computed as $\kappa(d)=\frac{-|Q|}{\nabla T(d)}\frac{1+\phi/2}{1-\phi}$; the factor $\frac{1+\phi/2}{1-\phi}$ is based on the Maxwell–Eucken relation~\cite{wang_new_2006} and is used to determine the conductivity of the dense (matrix) phase from heat-flux and temperature-gradient measurements performed on a porous phase. The Maxwell–Eucken relation is used under the assumption that the inside-pore thermal conductivity is negligible (zero) compared to the matrix conductivity. The XRD analyses in Fig 2 of Ref.~\cite{curtis2010postmortem}, and in Figs. 3,7 of Ref.~\cite{wereszczak_postmortem_2000}, show that the `postmortem' composition of refractory silica bricks fired for years in furnaces\cite{kasaihistory} is mainly tridymite in proximity of the cold end.
 Therefore, we extracted the conductivity from the brick zones close to the cold end, these are labeled with `8' and `9' in Ref.~\cite{Pierce_1936} and have an average porosity $\phi=0.245$. More precisely, the purple empty diamonds in Fig.~\ref{fig:k_main_text} correspond to the thermal conductivity of TOGA HP tridymite (9-inch uninsulated brick) at distances from the hot end ranging from 8.62 to 7.82 inches (corresponding to temperatures raging from 745 to 892 K).

\subsection{Vitreous silica} 
\label{sub:subsection_name}
Vitreous silica was simulated using the GAP potential as discussed in Ref.~\cite{simoncelli_thermal_2023} (the solid blue line in Fig.~\ref{fig:k_main_text} corresponds to the conductivity of the `192(D)' model studied with GAP in Ref.~\cite{simoncelli_thermal_2023}, which is in agreement with the larger `5184(G)' model discussed in the same reference). We also note that our theoretical predictions for the conductivity of vitreous silica are in agreement with the result obtained by a recent, independent work \cite{fiorentino_hydrodynamic_2023}. The analysis of the geometric disorder in Fig.~\ref{fig:k_main_text}\textbf{b} and Fig.~\ref{fig:geom_disorder_appendix} was performed using the larger `5184(G)' model discussed in Ref.~\cite{simoncelli_thermal_2023}.

\subsection{Raman simulation} 
\label{sub:raman_simulations}

The Raman intensity of the vibrational mode $s$ at wavevector $\bm{q}=\bm{0}$ is computed following Eq.~(1) in Ref.~\cite{PhysRevLett.90.036401}, and accounting for the total intensity \cite{PhysRevB.71.214307}
\begin{equation}
\begin{split}
    I_s(T)&=I_{s,\parallel}(T) +I_{s,\perp}(T) \\
    &\propto\Bigg[ \left|\bm{e}_{\parallel} \cdot \stackrel{\leftrightarrow}{\bm{A}}_{s} \cdot \bm{e}_{\parallel}\right|^2 +
\left|\bm{e}_{\parallel} \cdot \stackrel{\leftrightarrow}{\bm{A}}_{s} \cdot \bm{e}_{\perp}\right|^2\Bigg]
   \frac{(\omega_L{-}\omega_{s})^4}{\omega_{s}}[\bar{\tenscomp{N}}_{s}{+}1],
\label{eq:raman_intensity}
\end{split}
\end{equation}
where $\stackrel{\leftrightarrow}{\bm{A}}_{s}$ is the Raman tensor, $\bar{\tenscomp{N}}_{s}$ is the Bose-Einstein distribution at temperature $T$, $\hbar\omega_{s}$ is the energy of the vibration, and $\hbar\omega_L=18726.59$ cm$^{-1}$ is the Laser energy (corresponding to the wavelength of 534 nm used in the experiments, more on this later).
The propagation direction of the laser with respect to the crystal  is $\bm{k}_L = 0.65 \bm{a} - 4.7 \bm{b} + 0.715 \bm{c}$, where $\bm{a}$, $\bm{b}$, $\bm{c}$ are the direct lattice vectors of the structure discussed by Dollase \textit{et al.}\cite{Dollase76}.
$\bm{e}_{\parallel}$ is the  polarization of the incident light, $\bm{e}_{\perp}$ is the direction orthogonal to $\bm{e}_{\parallel}$ in the polarization plane.

The Raman intensities are evaluated using density-functional perturbation theory (DFPT) \cite{PhysRevLett.90.036401} as implemented in \texttt{Quantum ESPRESSO}\cite{giannozzi2017advanced,revmodphys.73.515}. The electronic structure is calculated using the GAP-relaxed structure and employing the Local Density Approximation (LDA) functional. The LDA functional is chosen accounting for its proven accuracy in describing the properties of silica polymorphs\cite{PhysRevB.79.064202,hay_dispersion_2015}, as well as the constraints imposed by the DFPT module of \texttt{Quantum ESPRESSO}. We use Optimized norm-conserving Vanderbilt pseudopotentials (ONCVPSP)\cite{PhysRevB.88.085117} from the \texttt{PseudoDojo} library\cite{van2018pseudodojo}.
Kinetic energy cutoffs of 90 and 360 Ry are used for the wave functions and the charge density, respectively; the Brillouin zone is integrated with a Monkhorst-Pack mesh of $3{\times}3{\times}2$ $\bm{k}$ points.

Once the Raman intensities~(\ref{eq:raman_intensity}) are evaluated, we compute the temperature-dependent Raman spectrum as discussed in the main text (see Ref.~\cite{PhysRevB.94.155435} for details),
 \begin{equation}
  I(\omega) = \sum_{s} I_s \frac{\frac{1}{2}(\Gamma^{\rm anh}_s+\Gamma^{\rm iso}_s+\Gamma_{\rm ins})}{(\omega-\omega_s)^2+\frac{1}{4}(\Gamma^{\rm anh}_s+\Gamma^{\rm iso}_s+\Gamma_{\rm ins})^2}
\label{eq:raman_intensity_omega}
\end{equation}
where $\Gamma^{\rm anh}_s$ are the temperature-dependent linewidth that account for third-order anharmonic scattering\cite{paulatto2013anharmonic} of the vibrational mode $s$ at $\bm{q}=\bm{0}$, $\Gamma^{\rm iso}_s$ is the temperature-independent linewidth accounting for scattering due to isotopic-mass disorder\cite{tamura1983isotope}, and $\Gamma_{\rm ins}=2.9\;$ cm$^{-1}$ is the instrumental linewidth (FWHM) of the Raman spectrometer used to perform the experiments in this work (details are reported later in the experimental section).

We note that the instrumental linewidth is smaller than the average energy-level spacing of MC tridymite, $\Delta\omega^{\rm av}_{\rm MC}\simeq 1300 {\rm cm}^{-1}/(72\times 3)\approx 6 {\rm cm}^{-1}$, thus the Raman experiments performed here have a resolution that is sufficiently high to distinguish individual vibrational modes.

\section{Experimental details} 
\label{sec:experimental_information}

\subsection{Sample preparation} 
\label{sub:sample_preparation}

MC tridymite was extracted from the fragment of Steinbach meteorite shown in Fig.~\ref{fig:raman}\textbf{a}; the dark regions in such figure correspond to tridymite, while the light-gray zones are identified as pyroxene, and the bright zones as Fe-Ni alloy.
A single tridymite grain, measuring approximately $1 {\times}1 {\times} 0.5$ mm (Fig.~\ref{fig:raman}\textbf{b} and \textbf{c}) was isolated from the meteorite fragment. 
Then, using a Triple Ion Beam Cutter system LEICA EM TIC 3X operating at -10$^\circ$C and 7kV, we prepared a flat surface of this isolated grain, and finally we embedded it in RS Pro Amber Epoxy (RS 199-1468). 
Environmental SEM imaging was performed using a TESCAN VEGA II VP-SEM/EDS fitted with a W filament operating with a voltage of 15 kV.

\subsection{X-ray diffraction measurements} 
\label{ssec:X-ray_experiments}
The X-ray (Mo K-alpha) diffraction (XRD) measurements were performed in the XRD platform of IMPMC (Paris) using a Rigaku/Agilent Xcalibur S 4-circle diffractometer. Four main crystalline domains were identified (relative proportions 0.43, 0.34, 0.15 and 0.08) using a single 80$^\circ$ phi scan (5s and 1$^\circ$ width/frame) corresponding to 867 independent reflections. The cell parameters corresponding to the main domain are a= 18.536(3) ~\AA, b= 5.0195(9) ~\AA, c= 23.825(4) ~\AA, beta= 105.780(19)$^\circ$, in excellent agreement with the meteoritic tridymite structure (space group $Cc$) determined by Dollase \textit{et al.}\cite{Dollase76}. This measurement was completed by a phi scan recorded on a limited 30$^\circ$ range on the sample mounted on the plot used for the other experimental measurements.
{We estimated that in the thermal conductivity experiment the temperature gradient interjected the largest  crystalline domain with relative direction $\bm{m}{=}0.65 \bm{a}{-}4.7\bm{b}{+}0.715\bm{c}$, where $\bm{a}$, $\bm{b}$, $\bm{c}$ are the direct lattice vectors of MC tridymite determined by Dollase \textit{et al.}\cite{Dollase76}. 
}

\subsection{Raman measurements} 
\label{ssec:raman_experiments}
Raman experiments (Fig.~\ref{fig:raman}\textbf{c}) were carried out at various temperatures using a confocal in-house optical setup using a Jobin-Yvon/Horiba HR-460 spectrometer equipped with a monochromator with 1500 grooves/mm and a Peltier-cooled Andor CCD. The instrumental linewidth (FWHM) was determined using a Neon lamp, and resulted 2.9 $\rm cm^{-1}$. A silver paste was used to mount the sample on the cold head of a liquid-helium flow cryostat kept under a high vacuum of 10$^{-6}$ mbar. Temperature was measured with a precision better than +/- 1K, using a CERNOX probe. Raman scattering was excited by a continuous argon ion laser emitting at wavelength 514.5 nm, and focused onto a spot of about 2 micrometers. The propagation direction of the laser with respect to the crystal was $k_L = 0.65 \bm{a} - 4.7 \bm{b} + 0.715 \bm{c}$, where $\bm{a}$, $\bm{b}$, $\bm{c}$ are the direct lattice vectors of the structure discussed by Dollase \textit{et al.}\cite{Dollase76}, was inferred from the results of the 30$^\circ$ XRD scan and comparison with the theoretical Raman spectra. Reported spectra correspond to the average of three spectra collected for 60 s in backscattering geometry with maximal 4 mW laser power on the sample.

\subsection{Thermal conductivity measurements} 
\label{ssec:Thermorflectance}
The thermal conductivity was measured via Modulated Thermoreflectance (MTR) technique. We employed the experimental configuration detailed in Ref.~\cite{ref1_MTR}. The setup involves two lasers: a green laser with a wavelength of 532 nm (Cobolt Samba 532nm 500mW) serving as the pump beam, and a blue laser with a wavelength of 488 nm (LBX-488 Oxxyus) serving as probe beam.

The pump beam is subjected to modulation using an acousto-optic modulator (MT80-A1 AA optoelectronic) operating at a selected frequency, within the range from 10 kHz to 1 MHz. An oscillating mirror is utilized to scan the pump beam across the surface of the sample, creating a heated region. The probe beam, on the other hand, remains stationary, reflecting off the device's surface onto a photodiode (New Focus model 1801 FS, 125 MHz). To detect the AC component of the reflected probe signal, a lock-in amplifier (SR 7280) is employed.

The pump beam typically delivers a few milliwatts and the probe beam at an approximate power of 150 microWatts. For precise alignment and focusing, a microscope objective (Olympus BX RFA) is used, allowing us to visualize the laser spots on the device. Both laser beams are assumed to exhibit Gaussian profiles, leading to an effective diameter of about $2.2 \mu$m.

The tridymite sample was placed in a Linkam stage positioned under the microscope, enabling reflectivity measurements and temperature control in the range from 120 K to room temperature. Due to the extremely low optical absorption of tridymite at 532 nm, a thin layer of gold (100 nm thick) is applied to cover the sample.

The signal measured in MTR experiments directly corresponds to changes in the optical reflection of the probe beam as a function of temperature variations within the region heated by the pump beam. The setup is optimized for measuring samples covered with gold, with the maximum $dR/dT$ observed at 488 nm. We systematically record the amplitude and phase of the probe signal while linearly scanning across the device surface at different distances from the pump-laser spot.

Recognizing the influence of both thickness and temperature on the thermal conductivity of gold ($\kappa_{\rm Au}$), we have refined its determination with precision under identical temperature conditions as those used in experiments involving MC tridymite and silica. This was accomplished by performing a MTR analysis on a 100 nm-thick gold film, which was deposited on a BK7 substrate whose thermal properties are well-known.  The $\kappa_{\rm Au}$ values obtained from this study align with our previous findings from similar experimental setups \cite{fretigny_analytical_2007}.

Data analysis is conducted through the solution of the three-dimensional model detailed in Refs.~\cite{ref1_MTR,ref2_MTR}. The model takes into account the heating effects induced by an intensity-modulated Gaussian laser beam, encompassing lateral and vertical heat diffusion. The amplitude and phase of the modulated temperature are determined by solving the heat-transport equation in three dimensions, utilizing standard Hankel transformations.

\noindent
{{\textbf{{Data availability}}}}\\   
The atomistic models of the silica polymorphs studied in this work are available on the Materials Cloud Archive.\\[3mm]

\noindent
{{\textbf{{Code availability}}}}\\ Quantum \textsc{espresso}\cite{giannozzi2017advanced} is available at \url{www.quantum-espresso.org}; the scripts related to the computation of the third order force constants using the finite-difference method are available at \url{bitbucket.org/sousaw/thirdorder}; 
\textsc{phonopy} and \textsc{phono3py} are available at \url{github.com/phonopy}.
The GAP potential for silica polymorphs is available at \cite{erhard2022machine}. \textsc{lammps}\cite{brown2011implementing} is available at \url{www.lammps.org} and the interface for \textsc{lammps} with the GAP potential\cite{Csanyi2007-py,GAP_potential,Kermode2020-wu}  is available at \url{github.com/libAtoms/QUIP}.\\

\noindent
{{\textbf{{Acknowledgements}}}}\\
M. S. acknowledges G. Csányi for useful discussions, as well as support from Gonville and Caius College, and from the SNSF project P500PT\_203178.
N. M. acknowledges funding from the Swiss National Science Foundation under the Sinergia grant no. 189924.
The numerical calculations have been performed on the Sulis Tier 2 HPC platform, funded by EPSRC Grant EP/T022108/1 and the HPC Midlands+consortium.
We acknowledge the Collection de Météorites du Muséum national d'Histoire naturelle (Paris) for providing the sample of Steinbach meteorite (Cohelper Id 157932).\\

\noindent
{{\textbf{{Author Contributions}}}}\\
M.S, N.M., and F.M. initiated the project. M.S. performed theoretical predictions that motivated the experimental measurements. E.B. coordinated the experimental work. K.B. performed the temperature-dependent Raman spectroscopy measurements, B.B and B.D. prepared and characterized the sample. D.F. and M.M. performed the thermal conductivity measurements. M.S. prepared the figures and wrote the first version of the manuscript; N.M, F.M., E.B., M.M., and D.F. contributed to its final version.\\

\noindent
{{\textbf{{Competing Interests}}}}\\
The authors declare no competing interests.

\end{document}